\documentclass[12pt] {article}
\usepackage{epsfig,graphicx,psfrag}

\normalsize
\newcommand{\gsim}{\hbox{ \raise3pt\hbox to 0pt{$>$}\raise-3pt\hbox{$\sim $} }}
\newcommand{\simgt}{\hbox{ \raise3pt\hbox to 0pt{$>$}\raise-3pt\hbox{$\sim$} }}
\newcommand{\lsim}{\hbox{ \raise3pt\hbox to 0pt{$<$}\raise-3pt\hbox{$\sim $} }}
\newcommand{\simlt}{\hbox{ \raise3pt\hbox to 0pt{$<$}\raise-3pt\hbox{$\sim$} }}
\newcommand{\be}{\begin{equation}}
\newcommand{\ee}{\end{equation}}
\newcommand{\bea}{\begin{eqnarray}}
\newcommand{\eea}{\end{eqnarray}}
\newcommand{\mathbold}[1]{\mbox{\boldmath $\bf#1$}}

\def\to{\rightarrow}
\newcommand{\figdir}{figs}
\begin{document}
%
\begin{titlepage}

\vspace*{2.0cm}

\begin{center}
{\Large \bf
How Well Can We Reconstruct the $\mathbold{\it{t\bar{t}}}$ System\\
Near its Threshold at Future $\mathbold{\it{{e^+e^-}}}$ Linear Colliders?
}
\end{center}

\vskip 1.5cm

\begin{center}
Katsumasa Ikematsu$^{a)}$
\footnote[1]{e-mail address: \texttt{ikematsu@post.kek.jp}},
Keisuke Fujii$^{b)}$
\footnote[2]{e-mail address: \texttt{fujiik@jlcuxf.kek.jp}},
Zenr\=o Hioki$^{c)}$
\footnote[3]{e-mail address: \texttt{hioki@ias.tokushima-u.ac.jp}},\\
Yukinari Sumino$^{d)}$
\footnote[4]{e-mail address: \texttt{sumino@tuhep.phys.tohoku.ac.jp}},
Tohru Takahashi$^{e)}$
\footnote[5]{e-mail address: \texttt{tohrut@hiroshima-u.ac.jp}}
\end{center}

\vskip 0.7cm

\begin{center}
a) Department of Physics, Hiroshima University,\\
Higashi Hiroshima 739-8526, Japan
\end{center}
\begin{center}
b) IPNS, KEK, Tsukuba 305-0801, Japan
\end{center}
\begin{center}
c) Institute of Theoretical Physics, University of Tokushima,\\
Tokushima 770-8502, Japan
\end{center}
\begin{center}
d) Department of Physics, Tohoku University,\\
Sendai 980-8578, Japan
\end{center}
\begin{center}
e) Graduate School of Advanced Sciences of Matter, Hiroshima University,\\
Higashi Hiroshima 739-8530, Japan
\end{center}

\vskip 0.7cm

\begin{abstract}
We developed a new method for full kinematical reconstruction
of the $t\bar{t}$ system near its threshold
at future linear $e^+e^-$ colliders.
In the core of the method lies likelihood fitting
which is designed to improve measurement accuracies
of the kinematical variables that specify the final states
resulting from $t\bar{t}$ decays.
The improvement is demonstrated by applying this method
to a Monte-Carlo $t\bar{t}$ sample generated
with various experimental effects including beamstrahlung,
finite acceptance and resolution of the detector system, etc.
In most cases the fit brings a broad non-Gaussian distribution
of a given kinematical variable to a nearly Gaussian shape,
thereby justifying phenomenological analyses
based on simple Gaussian smearing of parton-level momenta.
The standard deviations of the resultant distributions
of various kinematical variables are given in order to facilitate
such phenomenological analyses.
A possible application of the kinematical fitting method
and its expected impact are also discussed.
\end{abstract}
{\small {\it
}}

\end{titlepage}

%
\section{Introduction}
\label{Sec:introduction}

The discovery of the top quark~\cite{CDF-D0:1994} at Tevatron
has completed the standard-model (SM) list of matter fermions.
In spite of its subsequent studies thereat,
our knowledge on its properties is still far below the level
we reached for the lighter matter fermions.
A next-generation $e^+e^-$ linear collider such as JLC~\cite{JLC1},
having a facet as a top-quark factory,
is expected to allow us to measure top quark's properties
with unprecedented precision,
thereby improving this situation dramatically.
Such precision measurements may shed light on
the electroweak-symmetry-breaking mechanism
or hint beyond-the-SM physics or both.

Being aware of the opportunities provided by the linear collider,
a number of authors have so far performed interesting analyses
on the measurements of top quark properties
~\cite{Atwood:2000tu,ACFARep,Aguilar-Saavedra:2001rg,Abe:2001wn}.
They can be classified into two categories,
i.e., those near the $t\bar{t}$ threshold, mainly focused on
physics contained in the threshold enhancement factor,
and those in open-top region,
searching for anomalies in production and decay vertices,
both of which play important roles and complement each other.

In those analyses, feasibility studies on form factor measurements
have been done mainly in the open-top region
~\cite{Barklow:1995,Frey:1995ai,Iwasaki:2001ip}.
In the meantime, it has been conceived that, in view of
the energy upgrading scenario of the $e^+e^-$ linear collider,
measurements of top form factors in the $t\bar{t}$ threshold region
are also important.
The top quark physics is expected to commence in the threshold region
at the early stage of the collider operation
and full exploration of the machine potential in that phase
is crucial for the project design.
It has, therefore, been repeatedly stressed
that a realistic simulation study is in desperate need
to clarify feasibility of precision measurements of form factors
at the $t\bar{t}$ threshold.
Besides, form factor measurements in the $t\bar{t}$ threshold region
have some favorable features:
availability of well-controlled highly polarized top sample
~\cite{Gusken:1985nf,Peter:1997rk};
no need for transformation to $t$ or $\bar{t}$ rest frames
because both $t$ and $\bar{t}$ are nearly at rest;
as far as the decay form factor measurements of an on-shell top quark
are concerned, the center-of-mass energy does not matter.

In order to thoroughly carry out such analyses for real data,
we need a sophisticated method to kinematically reconstruct events
as efficiently and as precisely as possible.
This is, however, highly non-trivial in practice,
due to finite detector resolutions,
possible missing neutrinos in the final states,
and various background contributions.
Furthermore, care has to be taken when imposing a kinematical constraint
on the masses of the $t$ and $\bar{t}$ quarks
because they cannot be simultaneously on-shell below the threshold.
We thus need to further explore the potential
of the $e^+e^-$ linear collider and extend the past studies
~\cite{Barklow:1995,Frey:1995ai,Fujii:1993mk,Comas:1995rw}
to the threshold region, in order to make maximum use of
the linear collider's advantages:
clean experimental environment, well-defined initial state,
availability of highly polarized electron beam,
possibility of full parton-level reconstruction of final states, etc.

In this paper, we thus aim at developing an efficient method
for full kinematical reconstruction of the $t\bar{t}$ system
near its threshold in $e^+e^-$ annihilation
and clarifying the accuracy to which various observable will be measured.
We develop a likelihood fitting method which is especially designed
to improve measurement accuracies of kinematical variables
of the particles originating from the $t\bar{t}$ sample
in the threshold region.
Moreover, some of the analysis techniques developed here are expected
to be useful for the analyses in the open-top region.

Our study should also provide important information to the current line
of phenomenological studies on top quark physics at linear colliders.
In fact, there have been a number of theoretical studies on measurements
of the top-quark production and decay form factors
using  the $e^+e^- \to t\bar{t}$ process
~\cite{Atwood:2000tu,ACFARep,Aguilar-Saavedra:2001rg,Abe:2001wn}.
However, many of these analyses assumed
either the most optimistic case or the most conservative case
with respect to the kinematical reconstruction of event profiles.
In the former case, one assumes that the momenta of all the particles
(including $t$ and $W$) can be determined precisely,
while in the latter case, one uses only partial kinematical information,
e.g. the direction of $b$, and the energy and momentum of $\ell$.
In this work we will provide realistic values of resolutions
with which individual kinematical variables can be measured.

In Sec.~2 we briefly review our simulation framework.
Sec.~3 is devoted to top quark reconstruction
in the lepton-plus-4-jet mode, where two subsections
recapitulate basic strategy and procedure, respectively.
In Sec.~4 we explain our kinematical reconstruction
using a likelihood fitting method.
Then we discuss a possible application of this method
and its expected impact in Sec.~5.
Finally, Sec.~6 summarizes our results and concludes this paper.

%
\section{Framework of Analysis}
\label{Sec:framework}

For Monte-Carlo-simulation studies of $t\bar{t}$ productions and decays,
we developed an event generator that is now included
in \texttt{physsim-2001a}~\cite{PHYSSIM},
where the amplitude calculation and phase space integration are performed
with \texttt{HELAS/BASES}~\cite{Murayama:1992gi,Kawabata:1985yt}
and parton 4-momenta of an event are generated
by \texttt{SPRING}~\cite{Kawabata:1985yt}.
In the amplitude calculation, initial state radiation (ISR) as well as
$S$- and $P$-wave QCD corrections to the $t\bar{t}$ system
~\cite{Sumino:1992ai,Murayama:1992mg} are taken into account.
Parton showering and hadronization are carried out
using \texttt{JETSET 7.4}~\cite{Sjostrand:1993yb}
with final-state tau leptons treated by \texttt{TAUOLA}~\cite{Jadach:1993hs}
in order to handle their polarizations properly.

In this study, the top-quark (pole) mass is assumed to be 175~GeV
and the nominal center-of-mass energy is set
at 2~GeV-above the $1S$ resonance of the $t\bar{t}$ bound states.
This energy is known to be suitable for measurements of various properties
of the $t\bar{t}$ system at threshold~\cite{Fujii:1993mk}.
We will assume an electron-beam polarization of 80\% in what follows.
Effects of natural beam-energy spread and beamstrahlung are taken into account
according to the prescription given in \cite{Fujii:1993mk},
where the details of the beam parameters are also described.
We have assumed no crossing angle between the electron and the positron beams
and ignored the transverse component of the initial state radiation.
Consequently, the $t\bar{t}$ system in our Monte-Carlo sample has
no transverse momentum.
Under these conditions we expect 40k $t\bar{t}$ events for 100$fb^{-1}$.

The generated Monte-Carlo $t\bar{t}$ events were passed
to a detector simulator (\texttt{JSF Quick Simulator}~\cite{JSF})
which incorporates the ACFA-JLC study parameters (see Table.~\ref{det-param}).
The quick simulator created vertex-detector hits,
smeared charged-track parameters in the central tracker
with parameter correlation properly taken into account,
and simulated calorimeter signals as from individual segments,
thereby allowing realistic simulation of cluster overlapping.
It should also be noted that track-cluster matching was performed
to achieve the best energy-flow measurements.

\begin{table}[hbp]
 \begin{center}
   \begin{tabular}{|c||r|r|} \hline
     \textbf{Detector} & \textbf{Performance} & \textbf{Coverage} \\
     \hline
     \hline
     Vertex detector                                                      &
       $ \sigma_{\mathrm{b}}
         = 7.0 \oplus (20.0/p) \, / \, \sin^{3/2}\theta \; \mu\mathrm{m}$ &
       $ | \cos\theta | \leq 0.90 $ \\
     \hline
     Central drift chamber                                                &
       $ \sigma_{p_{T}}/p_{T}
         = 1.1 \times 10^{-4} p_{T} \> \oplus \> 0.1 \; \% $              &
       $ | \cos\theta | \leq 0.95 $ \\
     \hline
     EM calorimeter                                                       &
       $ \sigma_{E}/E
         = 15 \; \% \> / \sqrt{E} \> \oplus 1 \; \% $                     &
       $ | \cos\theta | \leq 0.90 $ \\
     \hline
     Hadron calorimeter                                                   &
       $ \sigma_{E}/E
         = 40 \; \% \> / \sqrt{E} \> \oplus 2 \; \% $                     &
       $ | \cos\theta | \leq 0.90 $ \\
     \hline
   \end{tabular}
 \end{center}
 \caption[]{
       ACFA study parameters of the JLC detector, where
       $p$, $p_{T}$, and $E$ are measured in units of GeV.
       }
 \label{det-param}
\end{table}%

%
\section{Event Selection}

\subsection{Basic Reconstruction Strategy}

Since the top quark decays almost 100\% into a $b$ quark and a $W$ boson,
the signature of a $t\bar{t}$ production is two $b$ jets and two $W$ bosons
in the final state.
These $W$ bosons decay subsequently either
leptonically into a lepton plus a neutrino or hadronically into two jets.
According to how the $W$ bosons decay, therefore,
there will be three modes of final states:
(1) six jets, where both of the $W$'s decay hadronically,
(2) one lepton plus four jets, where one of the $W$'s decays
leptonically and the other hadronically, and
(3) two leptons plus two jets, where both of the $W$'s decay leptonically.

In order to reconstruct the momentum vector of the top quark,
we will use the lepton-plus-4-jet mode,
for which we can reconstruct the $t$($\bar{t}$)-quark momentum
as the momentum sum of the $b$($\bar{b}$) jet
and the two jets from the hadronically-decayed $W^+$($W^-$),
while we can tell the charge of the hadronically-decayed $W$
from the charge of the lepton.

In the lepton-plus-4-jet mode, two of the four jets are
$b$($\bar{b}$) jets directly from the $t$($\bar{t}$) quarks,
while the other two are from the $W$ boson that decayed hadronically.
Therefore, if one can identify the $b$ and $\bar{b}$ jets,
remaining two jets can be uniquely assigned as decay products of the $W$ boson.
The other $W$ boson can be reconstructed from the lepton
and the neutrino indirectly detected as a missing momentum.
Remaining task is then to decide which $b$($\bar{b}$) jet
to attach to which $W$-boson candidate,
in order to form $t$($\bar t$) quarks.
Since the $t$($\bar t$) quarks are virtually at rest near the threshold,
a $b$($\bar b$) jet and the corresponding $W$ boson fly
in the opposite directions.
We can thus choose the correct combination by requiring the $b$($\bar b$) jet
and the $W$ boson be approximately back-to-back.

In reality, however, $b$($\bar b$)-quark tagging is not perfect
and can be performed only with some finite efficiency and purity:
there could be more than two $b$($\bar{b}$)-jet candidates in a single event.
In addition, $b$ and $\bar{b}$ quarks can be emitted in the same direction.
In such a case, a wrong combination could accidentally satisfy
the back-to-back condition.
These facts sometimes prevent us from uniquely assigning each jet
to its corresponding parton, resulting in multiple solutions
for a single event.
Moreover, the leptonically-decayed $W$ is poorly reconstructed in practice,
since the neutrino momentum is strongly affected by ISR, beamstrahlung,
as well as other possible neutrinos emitted from the $b$($\bar{b}$) jets.
In order to overcome these difficulties,
we will need some sophisticated method.
We defer discussion of such a method to the next section and examine here
the extent to which the aforementioned basic reconstruction strategy works.

\subsection{Event Selection Procedure}

\begin{figure}[htbp]
 \begin{center}
   \includegraphics[height=6cm,clip]{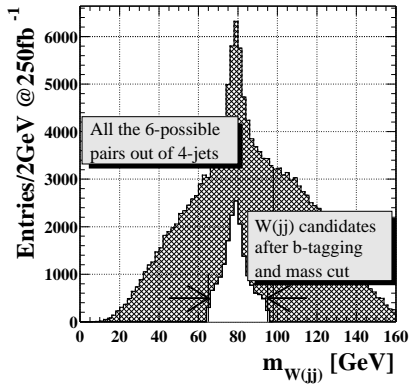}
 \end{center}
 \caption{
       Invariant mass distributions of the 2-jet systems
       reconstructed as $W$-boson candidates.
       Hatched and solid histograms correspond to
       before and after double $b$-tagging, respectively.
       The locations of the $W$ mass cuts are indicated with arrows.}
 \label{M2j}
\end{figure}

The lepton-plus-4-jet-mode selection started with demanding
an energetic isolated lepton:
$E_\ell > 18~{\rm GeV}$ and $E_{14^\circ {\rm cone}} < 18~{\rm GeV}$,
where $E_\ell$ is the lepton's energy
and $E_{14^\circ {\rm cone}}$ is the energy sum of particles
within a cone with a half angle of $14^\circ$ around the lepton direction
excluding the lepton itself.\footnote{
The $E_\ell$ cut was chosen to be the kinematical limit for the lepton
from the $W \to \ell\nu$ decay. On the other hand, the cone-energy cut
was optimized to achieve high purity, while keeping reasonable efficiency.}
When such a lepton was found,
the rest of the final-state particles was forced clustering to four jets,
using the Durham clustering algorithm~\cite{Catani:1991hj}.
Two-jet invariant mass was then calculated for
each of the six possible combinations
and checked if it was between 65~GeV and 95~GeV,
in order to select a jet pair which was consistent
with that coming from a $W$-boson decay.
For such a jet pair the remaining two jets, at the same time,
had to be identified as $b$($\bar{b}$) jets,
using flavor tagging based on the impact parameter method.
The hatched histogram in Fig.~\ref{M2j} is the 2-jet invariant mass
distribution of all the possible pairs out of the four jets,
while the solid histogram being that with the $b$-tagging.
It is seen that this procedure dramatically improved
the purity of the $W$ boson sample.
It should also be stressed that these selection
criteria are very effective to suppress background processes
such as $e^{+}e^{-} \to W^{+}W^{-}$
and provide us with an essentially background-free $t\bar{t}$ event sample.

\begin{figure}[hbtp]
 \begin{center}
   \includegraphics[height=6cm,clip]{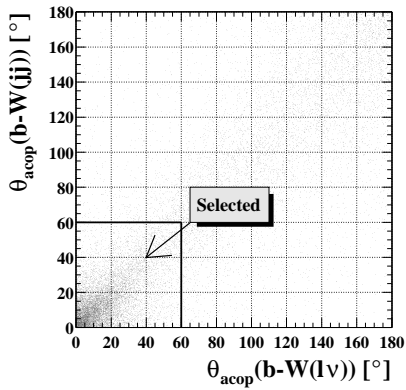}
 \end{center}
 \caption{
       Scatter plot of the acoplanarity angles corresponding to
       two $b$-$W$ systems, where horizontal and vertical axes are angles
       of $b$-$W_{\ell\nu}$ and $b$-$W_{2-jet}$ systems, respectively.}
 \label{AcopbW}
\end{figure}

The remaining task is to decide which $b$($\bar{b}$) jet
to associate with which $W$ candidate.
For a $b$($\bar{b}$)-jet candidate,
the right $W$ boson partner was selected
by requiring the back-to-back condition as described above.
Fig.~\ref{AcopbW} is a scatter plot of the acoplanarity angles
of the two possible $b$-$W$ systems
where horizontal and vertical axes are the angles
of $b$-$W_{\ell\nu}$ and $b$-$W_{2-jet}$ system, respectively.
$b$-$W$ pairs having $\theta_{acop(b-W)} \leq 60^\circ$ was
regarded as daughters of the $t$($\bar t$) quarks.

The selection efficiency after all of these cuts was found to be 15\%
including the branching fraction to the lepton-plus-4-jet mode of 29\%.

%
\section{Kinematical Fit}

The event selection described above yields a very clean $t\bar{t}$ sample.
As noted above, however, the sample is still subject
to combinatorial backgrounds,
if we are to fully reconstruct the final state by assigning each jet
to a corresponding decay daughter of the $t$ or $\bar{t}$ quark.
We thus need a well-defined criterion to select the best
from possible multiple solutions.
It is also desirable to improve the measurement accuracies
of those kinematical variables which are suffering from effects of
missing neutrinos (such variables include momenta of $b$, $\bar{b}$
or the neutrino from a $W$ itself).

The $t\bar{t}$ system produced via $e^+e^-$ annihilation
is a heavily constrained system: there are many mass constraints
in addition to the usual 4-momentum conservation.
At $e^+e^-$ linear colliders,
thanks to their well-defined initial state and the clean environment,
we can make full use of these constraints and perform a kinematical fit
to select the best solution and to improve the measurement accuracies
of the kinematical variables of the final-state partons.

\subsection{Parameters, Constraints, and Likelihood Function}

\begin{figure}[htbp]
  \begin{center}
    \includegraphics[height=5cm,clip]{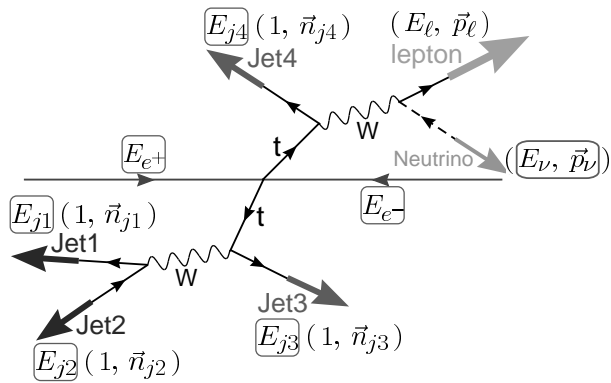}
  \end{center}
  \caption{
	Schematic diagram showing parameters and constraints
	relevant to the kinematical fit described in the text.
	The boxed parameters are unknown and to be determined by the fit.}
  \label{kine-fit}
\end{figure}

For the lepton-plus-4-jet final state,
there are 10 unknown parameters to be determined by the fit:
the energies of four jets,
the 4-momentum of the neutrino from the leptonically-decayed $W$ boson,
and the energies of the initial-state electron and positron,
provided that the jet directions as output from the jet finder
are accurate enough,
the error in the 4-momentum measurement of the lepton
from the leptonically-decayed $W$ can be ignored,
and that the transverse momenta of the initial-state electron and positron
after beamstrahlung or initial-state radiation or both are
either negligible or known from a low angle $e^+/e^-$ detector system
\footnote{ In addition, there will be some finite transverse momenta
due to a finite crossing angle of the two beams.
These transverse momenta are, however,
known and can be easily incorporated into the fit.}(see Fig.~\ref{kine-fit}).

The requirements of 4-momentum conservation and the massless constraint
for the neutrino from the leptonically-decayed $W$
reduce the number of free parameters to 5.
We choose, as these free parameters,
the energies of the four jets and the initial longitudinal momentum
(the difference of the energies of the initial-state electron and positron).

These five unknown parameters can be determined by
maximizing the following likelihood function:
\begin{equation}
  L = ( \prod_{f=1}^{4} P^f_{E_{f}}(E_{f}^{measured}, E_{f}))
      \cdot P_{\Gamma_{W^{+}}}
      \cdot P_{\Gamma_{W^{-}}}
      \cdot P_{\Gamma_{t\bar{t}}}
      \cdot P_{\sqrt{s}},
\end{equation}
where $P^f_{E_{f}}$ is a resolution function for jet $f$
and is Gaussian for $f=1$ and $2$ (jets from the hadronically-decayed $W$)
as given by the detector energy resolution.
For $f=3$ and $4$ (jets from the $b$ and $\bar{b}$ quarks)
the resolution function is the same Gaussian convoluted
with the missing energy spectrum due to possible neutrino emissions.
For the two $W$ bosons in the final state,
we use a Breit-Wigner function $P_{\Gamma_{W}}$
instead of $\delta$-function-like mass constraints.
$P_{\sqrt{s}}$ is a weight function coming from ISR and beamstrahlung effects.
This distribution was calculated as a differential cross section
as a function of the energies of initial-state electron and positron,
taking into account the $t\bar{t}$ threshold correction
as described in Sec.~\ref{Sec:framework}.

\begin{figure}[htbp]
  \begin{center}
    \includegraphics[height=5cm,clip]{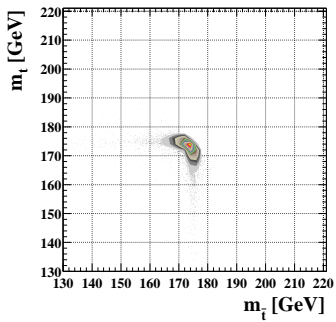}
  \end{center}
  \caption{$P_{\Gamma_{t\bar{t}}}$ distribution below $t\bar{t}$ threshold.}
  \label{Mttbar-gen}
\end{figure}

The remaining factor, $P_{\Gamma_{t\bar{t}}}$,
controls the mass distribution of the $t$ and $\bar{t}$ quarks
and has been introduced to take into account the kinematical constraint
that the $t$ and $\bar{t}$ cannot be simultaneously on-shell below threshold
(see Fig.~\ref{Mttbar-gen} which shows $P_{\Gamma_{t\bar{t}}}$ distribution
below $t\bar{t}$ threshold).
$P_{\Gamma_{t\bar{t}}}$ distribution is a dynamics-independent factor
which is extracted from the theoretical formula
for the threshold cross section.

\subsection{Results}

We performed the maximum likelihood fit for the selected sample.
The maximum likelihood fit provided us with a well-defined
clear-cut criterion to select the best solution,
when there were multiple possible solutions for a single event:
we should select the one with the highest likelihood.

\begin{figure}[htbp]
  \begin{center}
    \includegraphics[height=5cm,clip]{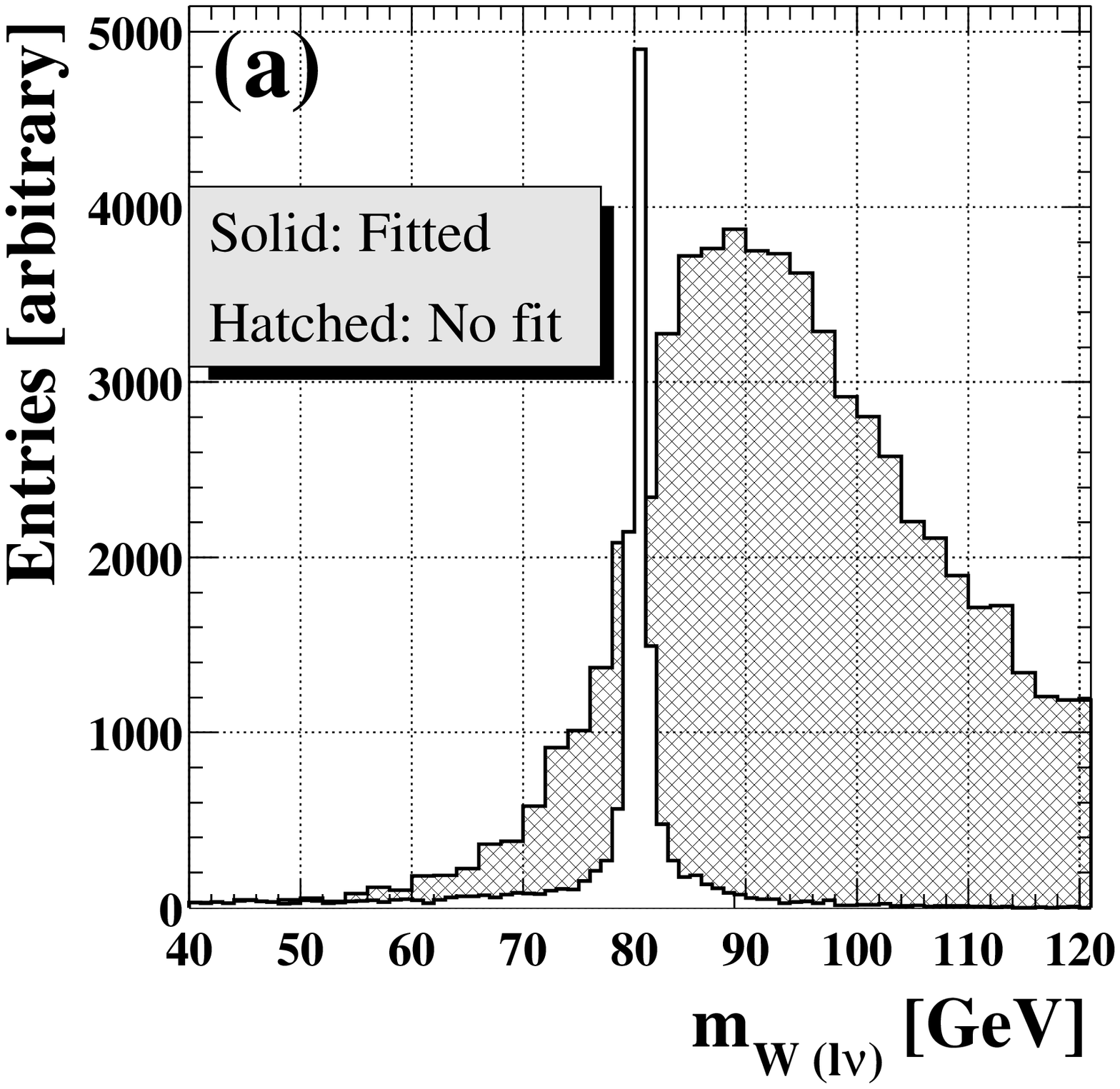}
    \includegraphics[height=5cm,clip]{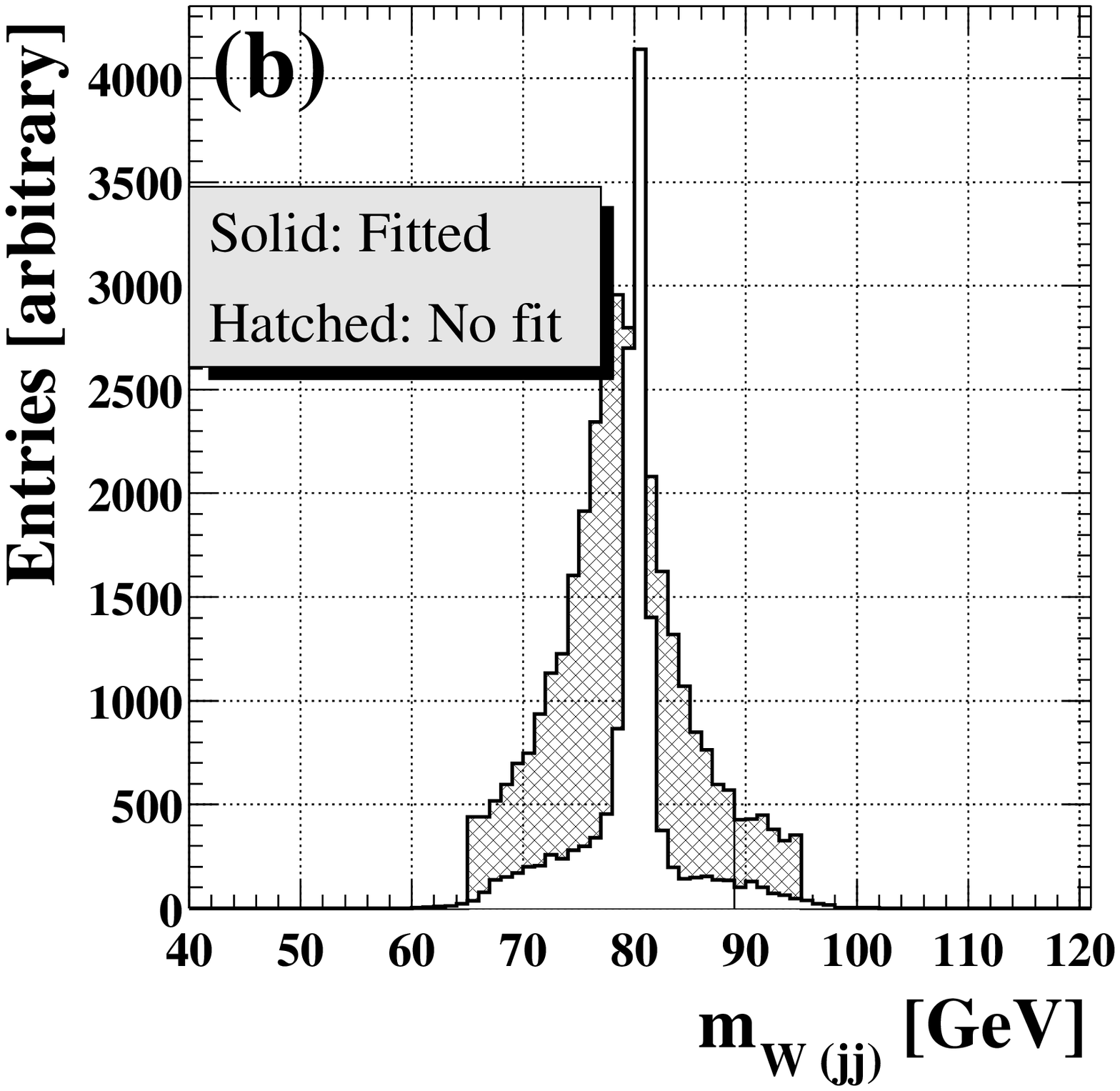}
  \end{center}
  \caption{
	Reconstructed $W$ mass distributions for
	(a) leptonically and (b) hadronically-decayed $W$ bosons,
	before (hatched) and after (solid) the kinematical fitting.
	Note that the vertical scale for the hatched area is enhanced
	by a factor of five for illustration purpose.}
  \label{Mww}
\end{figure}

Figs.~\ref{Mww}-a) and -b) are the reconstructed $W$ mass
distributions for the leptonically and hadronically-decayed
$W$ bosons, respectively, before (hatched) and after (solid)
the kinematical fit.
The figures demonstrate that the Breit-Wigner factors
($P_{\Gamma_{W^{\pm}}}$) in the likelihood function
properly constrain the $W$ masses as intended.

\begin{figure}[hptb]
  \begin{center}
    \includegraphics[height=5cm,clip]{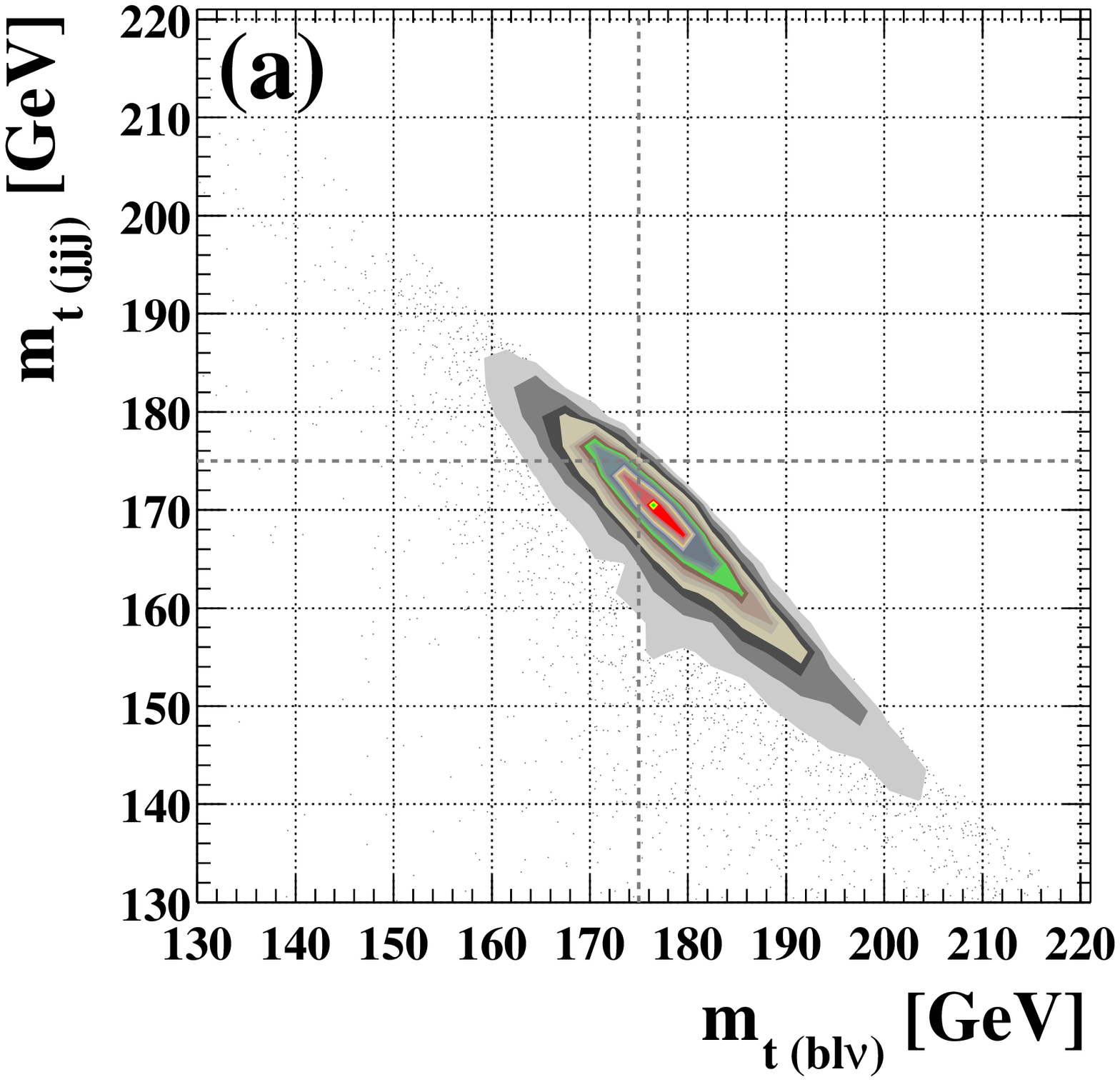}
    \includegraphics[height=5cm,clip]{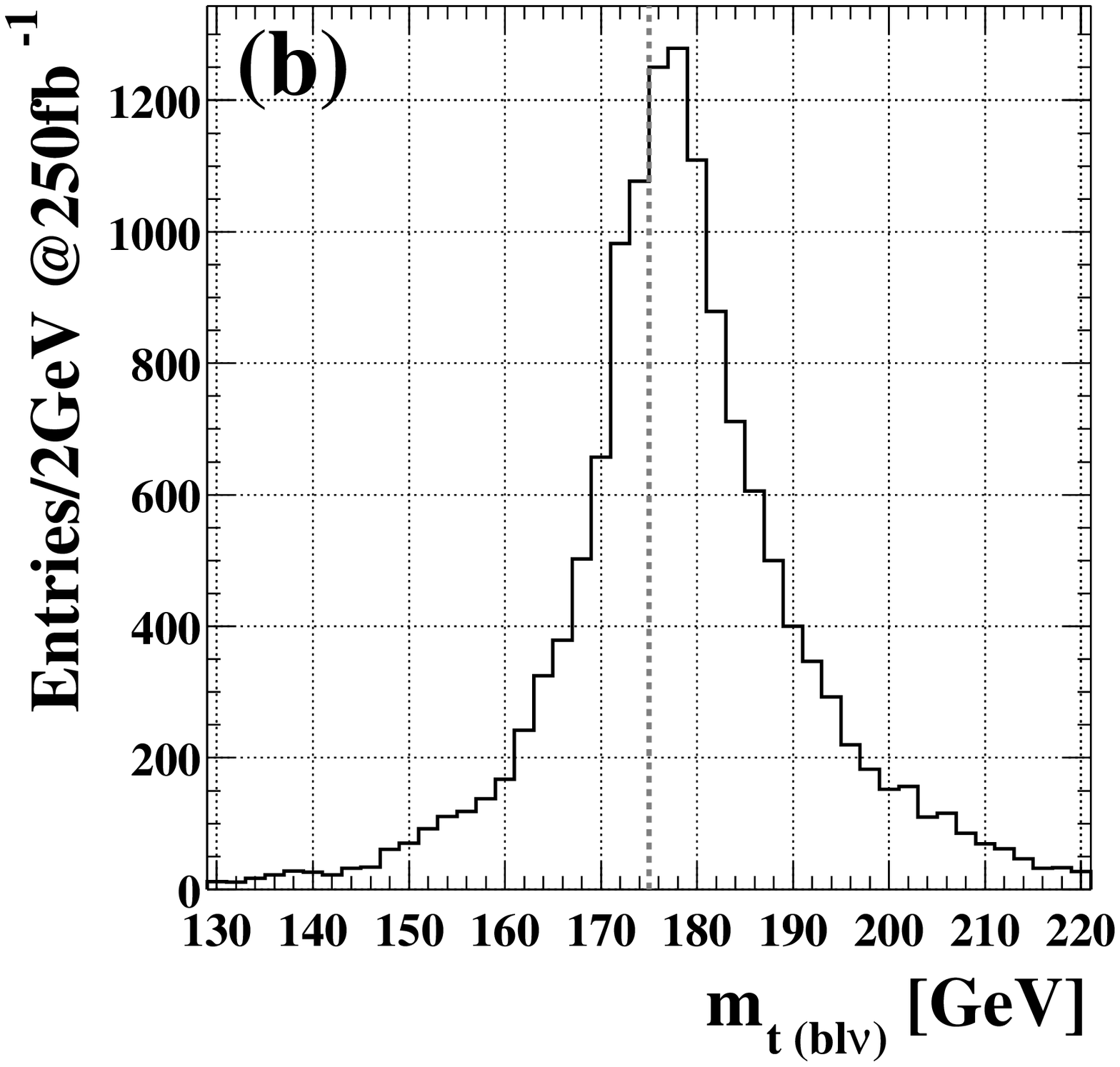}
    \includegraphics[height=5cm,clip]{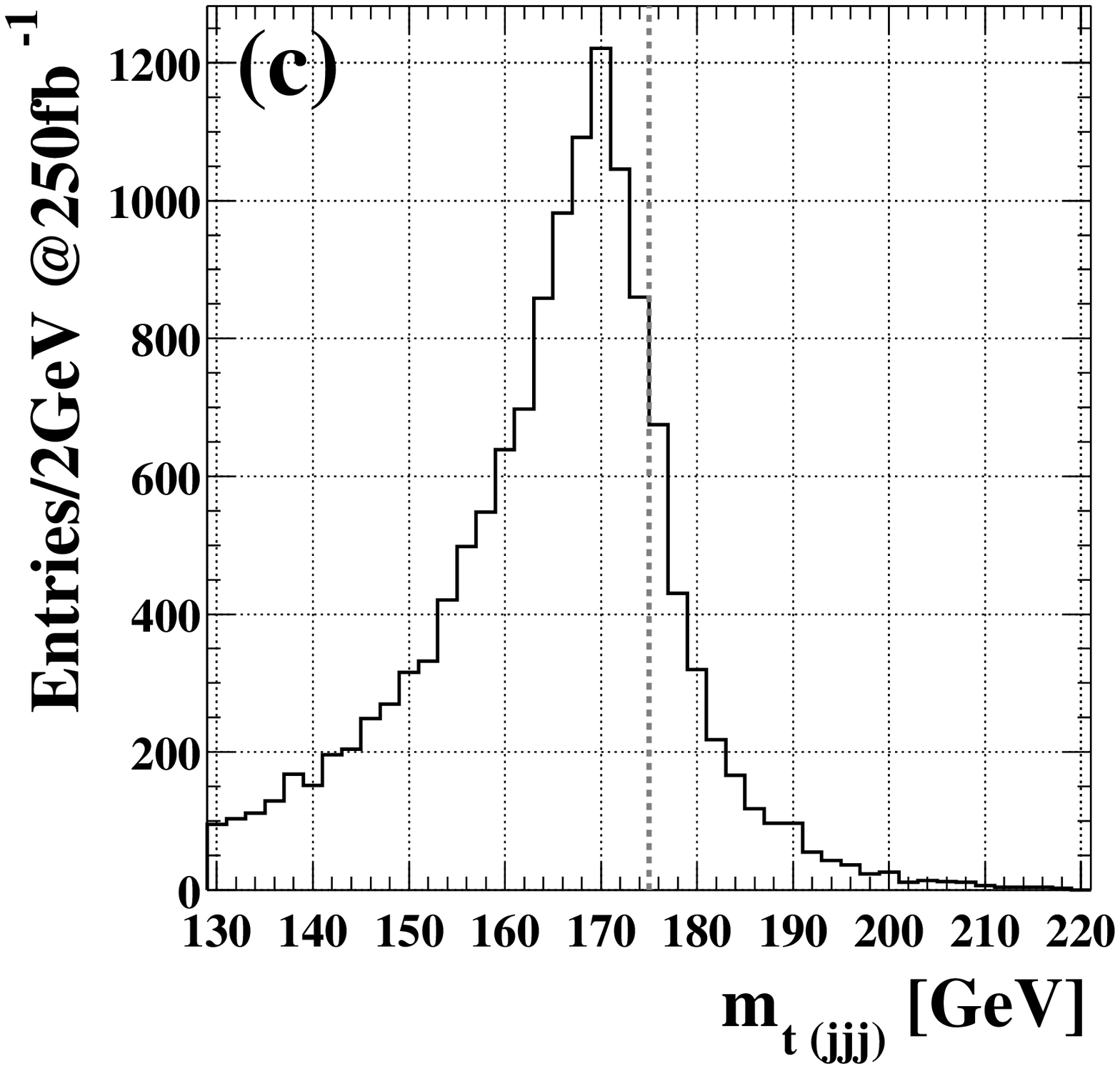}
    \\
    \includegraphics[height=5cm,clip]{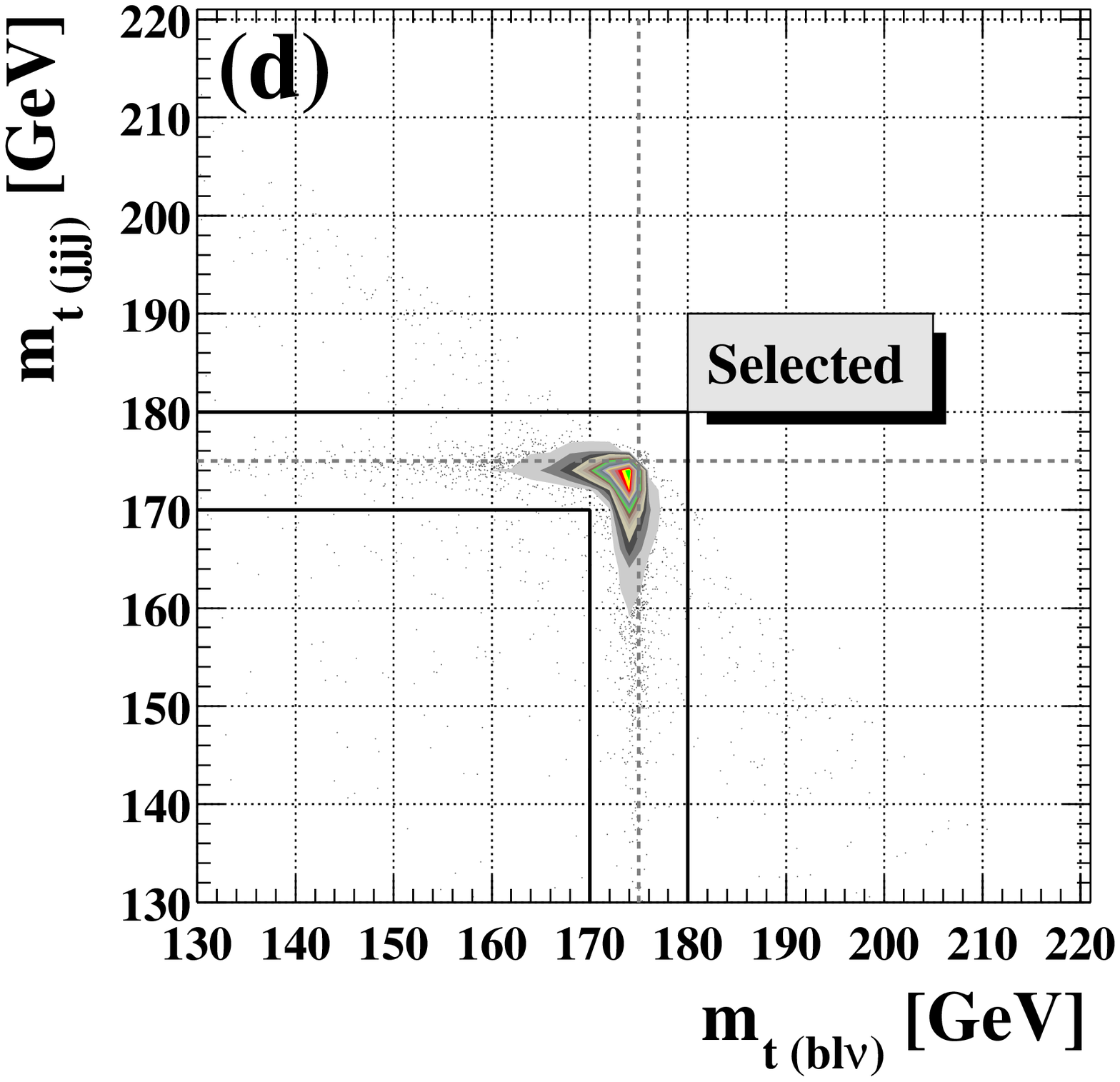}
    \includegraphics[height=5cm,clip]{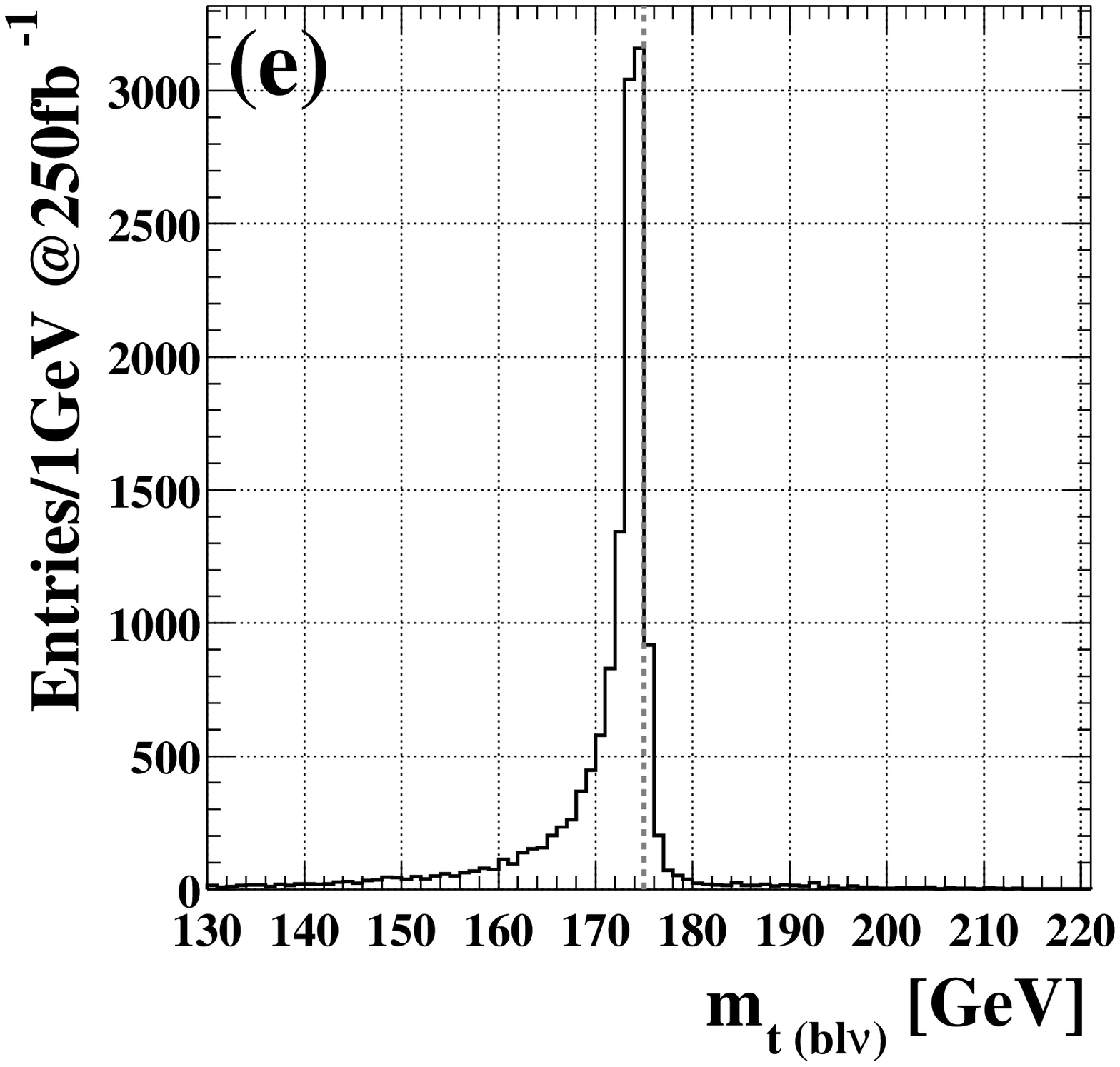}
    \includegraphics[height=5cm,clip]{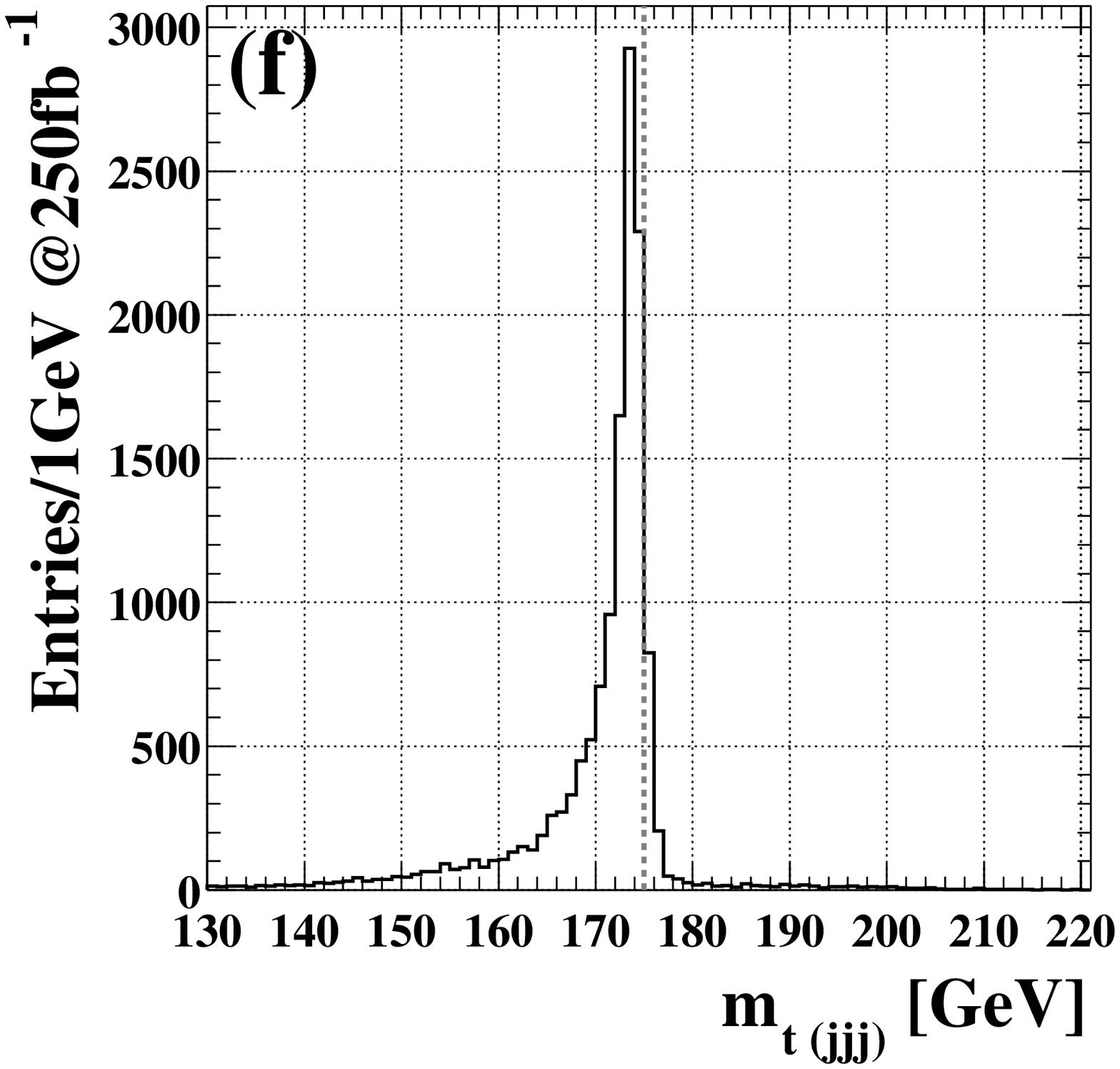}
    \\
    \includegraphics[height=5cm,clip]{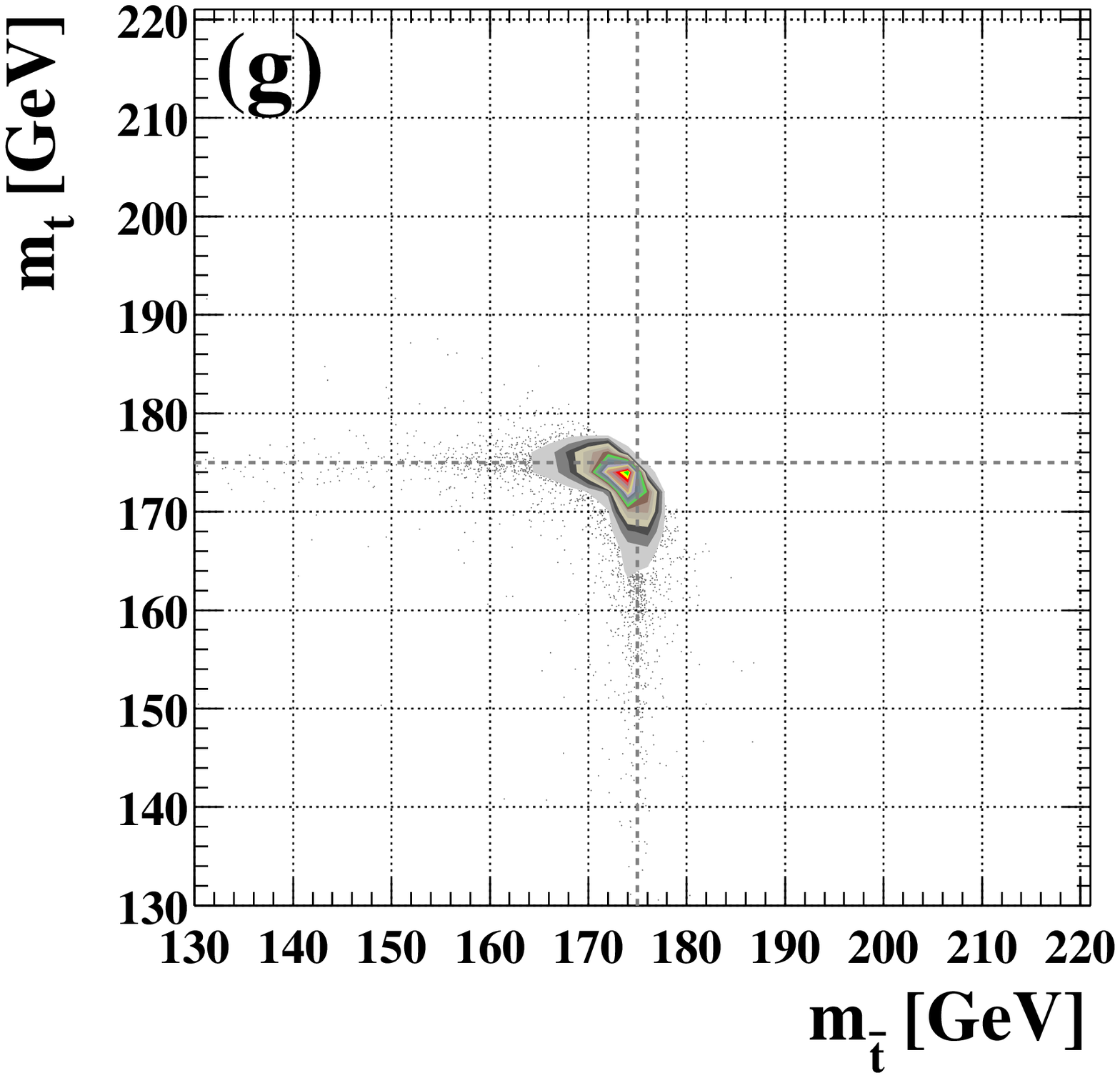}
    \includegraphics[height=5cm,clip]{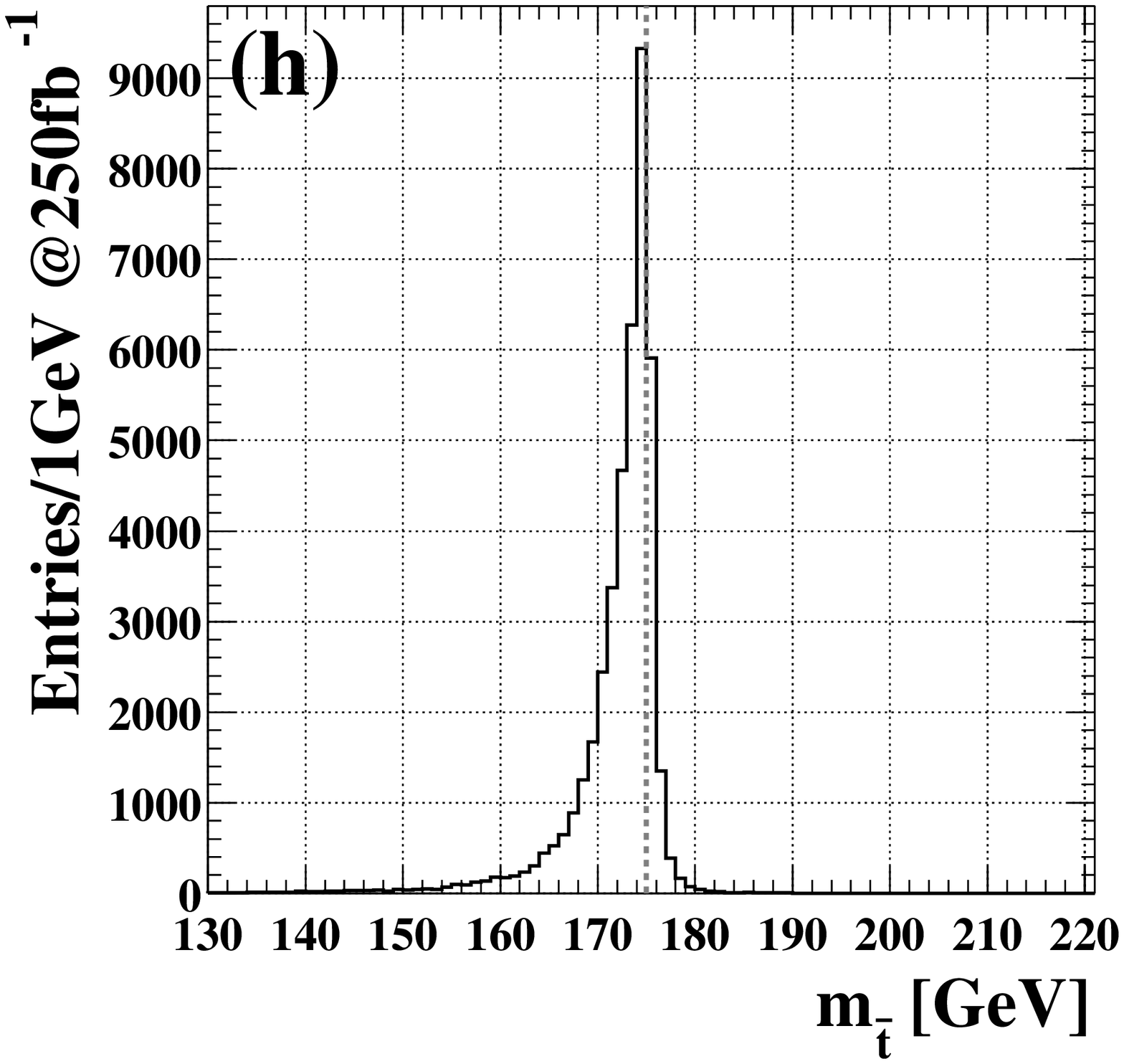}
    \includegraphics[height=5cm,clip]{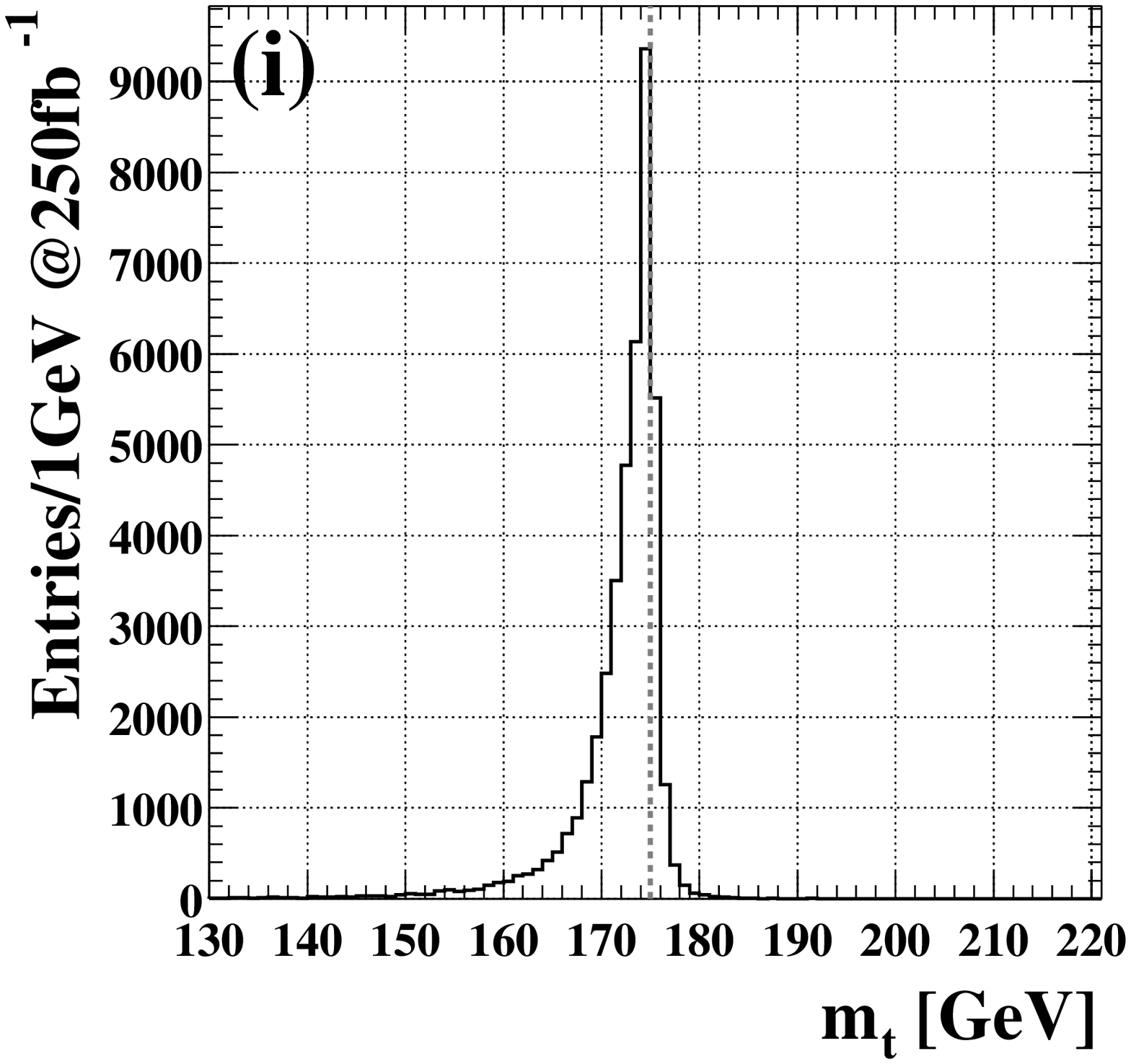}
  \end{center}
  \caption{
	Scatter plot of the reconstructed $t$($\bar{t}$) mass
	from 3 jets versus that from a lepton plus a $b$-jet
	(a) before the kinematical fit,
	together with (b) its horizontal/$b\ell\nu$
	and (c) vertical/3-jet projections.
	(d) through (f) are similar plots after the kinematical
	fit and (g) though (i) are corresponding plots for
	generated values.}
  \label{Mttbar}
\end{figure}

Fig.~\ref{Mttbar}-a) plots the reconstructed mass for the $t$($\bar{t}$)
decayed into 3 jets against that of the $\bar{t}$($t$)
decayed into a lepton plus a $b$ jet, before the kinematical fit.
The strong negative correlation is due to the fact
that the neutrino from the leptonically-decayed $W$ is reconstructed
as the total missing momentum.
Figs.~\ref{Mttbar}-b) and -c) are the projections of Fig.~\ref{Mttbar}-a)
to the horizontal and vertical axes, respectively,
showing systematic shifts of the peak positions.\footnote{
This is in contrast with the result in~\cite{Fujii:1993mk},
where a quite tight set of cuts
was imposed upon the reconstructed $W$ and $t$ masses,
and consequently their peak shifts were less apparent
at the cost of significant loss of usable events.
The goal of this study is to establish an analysis procedure
to restore those events which would have been lost,
by relaxing the tight cuts while keeping reasonable accuracy for
event reconstruction.}
Figs.~\ref{Mttbar}-d) through -f) are similar plots to
Figs.~\ref{Mttbar}-a) through -c) after the kinematical fitting,
while Figs.~\ref{Mttbar}-g) through -i) are corresponding
distributions of generated values (Monte-Carlo truth).
The kinematical fit sent most of the events to the L-shaped region
indicated in Fig.~\ref{Mttbar}-d), as it should,
and made the distribution look like
the generated distribution shown in Fig.~\ref{Mttbar}-g).
Consequently, the peak shifts observed in the Figs.~\ref{Mttbar}-b) and -c)
have been corrected as seen in Figs.~\ref{Mttbar}-e) and -f).
There are, however, still some small fraction of events
left along the minus $45^\circ$ line.
These events were so poorly measured that it was impossible to restore.
The cut (angled region) indicated in Fig.~\ref{Mttbar}-d)
allowed us to remove them without introducing any strong bias
on the reconstruction of the kinematical variables.

\begin{figure}[hptb]
  \begin{center}
    \includegraphics[height=5cm,clip]{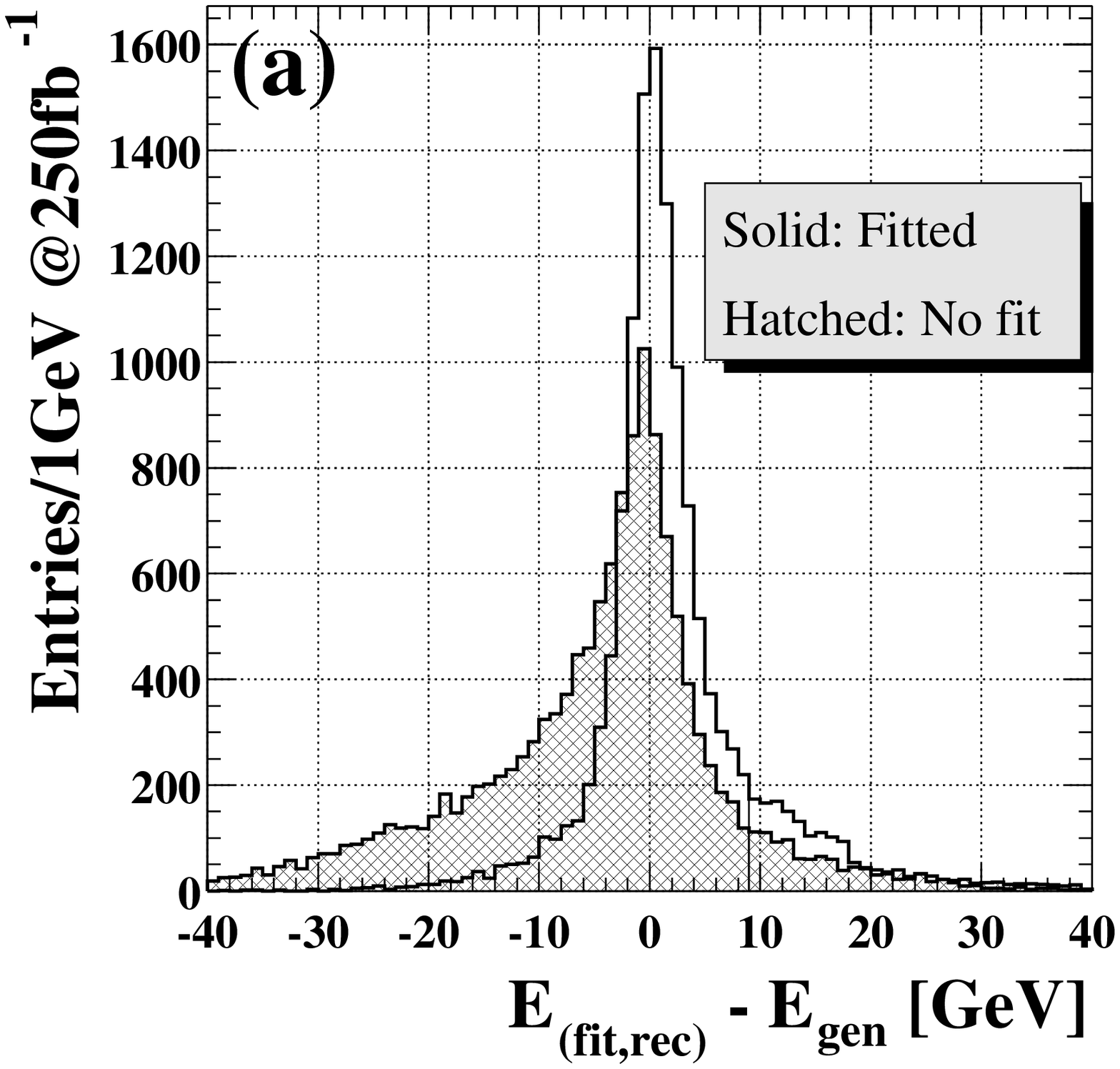}
    \includegraphics[height=5cm,clip]{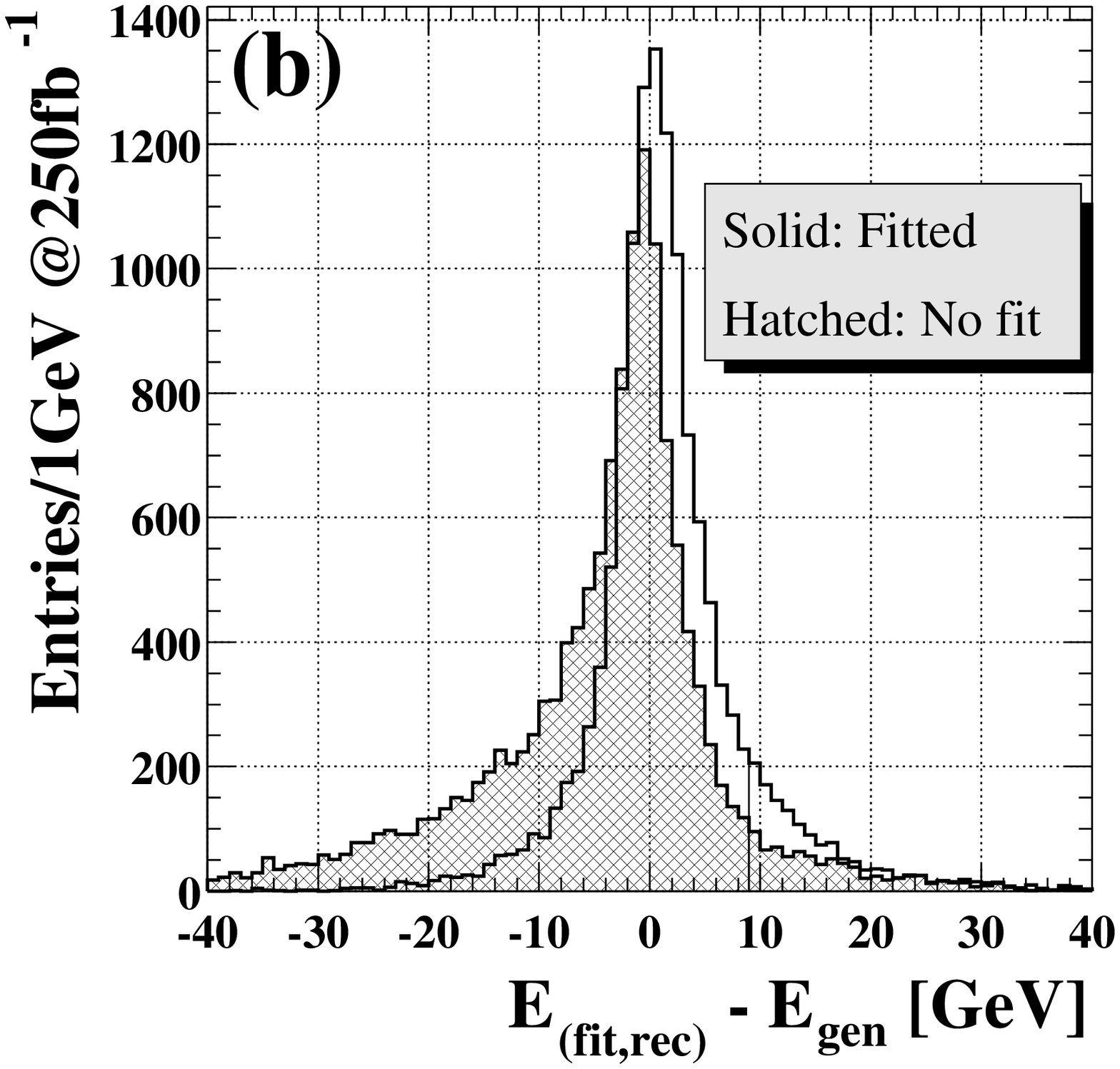}
    \\
    \includegraphics[height=5cm,clip]{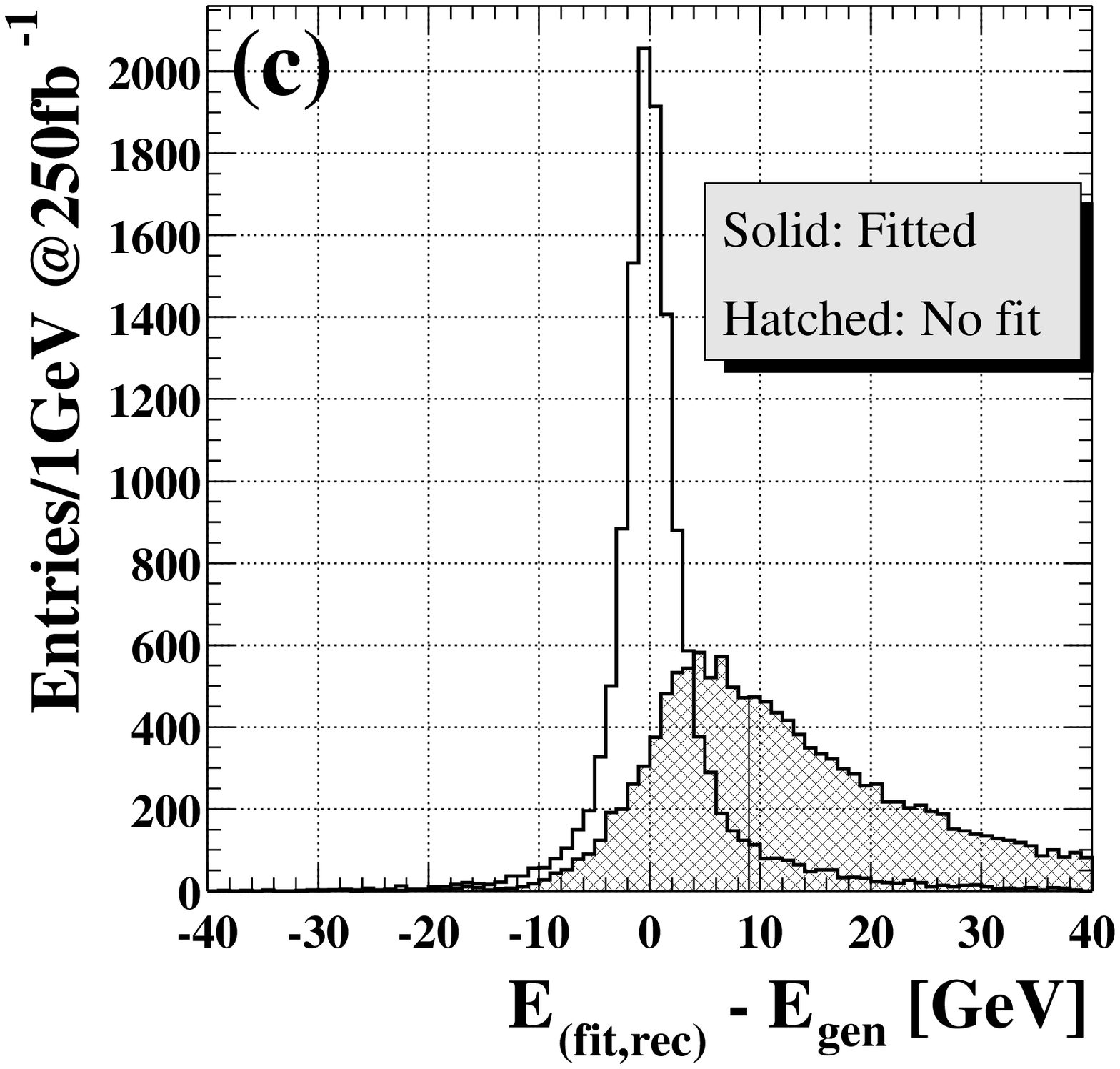}
    \includegraphics[height=5cm,clip]{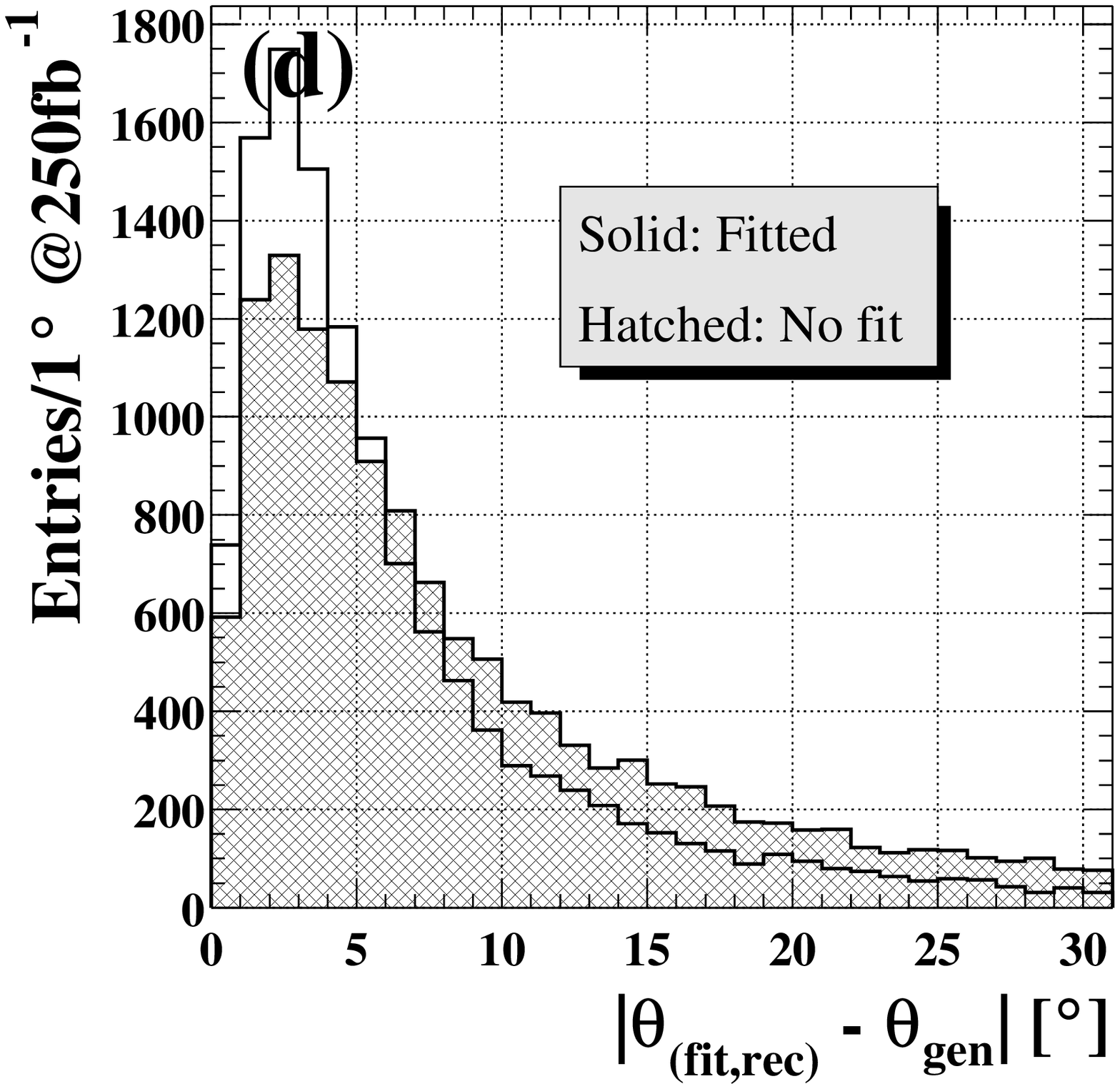}
  \end{center}
  \caption{
	Distributions of the difference of the reconstructed and generated
	energies of the $b$ or $\bar{b}$ jet attached to
	(a) leptonically-decayed and (b) hadronically-decayed $W$ bosons,
	together with distribution of the difference of
	the reconstructed and generated (c) energies
	and (d) directions of the direct neutrino
	from the leptonically-decayed $W$,
	before (hatched) and after (solid) the kinematical fit.}
  \label{Eresbnu}
\end{figure}

Now the question is how the above constraints improve the parameters
of the fit such as the energies of $b$ and $\bar{b}$ jets,
the direction and the magnitude of the missing neutrino
from the leptonically-decayed $W$,
on which we expect significant influences.
Figs.~\ref{Eresbnu}-a) and -b) plot the difference
between the reconstructed and the generated energies
of the $b$ ($\bar{b}$) quark attached to the leptonically-decayed $W$
and that of the $\bar{b}$ ($b$) attached to the hadronically-decayed $W$,
respectively, before (hatched) and after (solid) the kinematical fit.
The plots demonstrate that the kinematical constraints recover
the energies carried away by neutrinos from the $b$ or $\bar{b}$ decays.
The kinematical fit brings broad non-Gaussian distributions
into nearly Gaussian shapes.
The standard deviations of the $b$ or $\bar{b}$ jet energy distributions
are approximately 3.5~GeV after the kinematical fit.

The improvement is more dramatic for the direct neutrino
from the leptonically-decayed $W$, which is reconstructed
as the total missing momentum; see Figs.~\ref{Eresbnu}-c) and -d)
which show distributions of the difference of
the reconstructed and generated neutrino energies ($\Delta E_\nu$)
and directions ($\Delta \theta_\nu$).
Again the kinematical fit makes the broad and skewed distribution
of neutrino energies into a nearly Gaussian shape
with a standard deviation of approximately 2.5~GeV.
The fit also improves the angular resolution
as shown in Fig.~\ref{Eresbnu}-d).
The resultant angular resolution is $\sigma_{\theta} = 2.9^\circ$,
which was obtained by fitting
$\rm{N_0} \theta \exp(- \theta^2 / 2\sigma_\theta^2)$ to the distribution.

\begin{figure}[hptb]
  \begin{center}
    \includegraphics[height=5cm,clip]{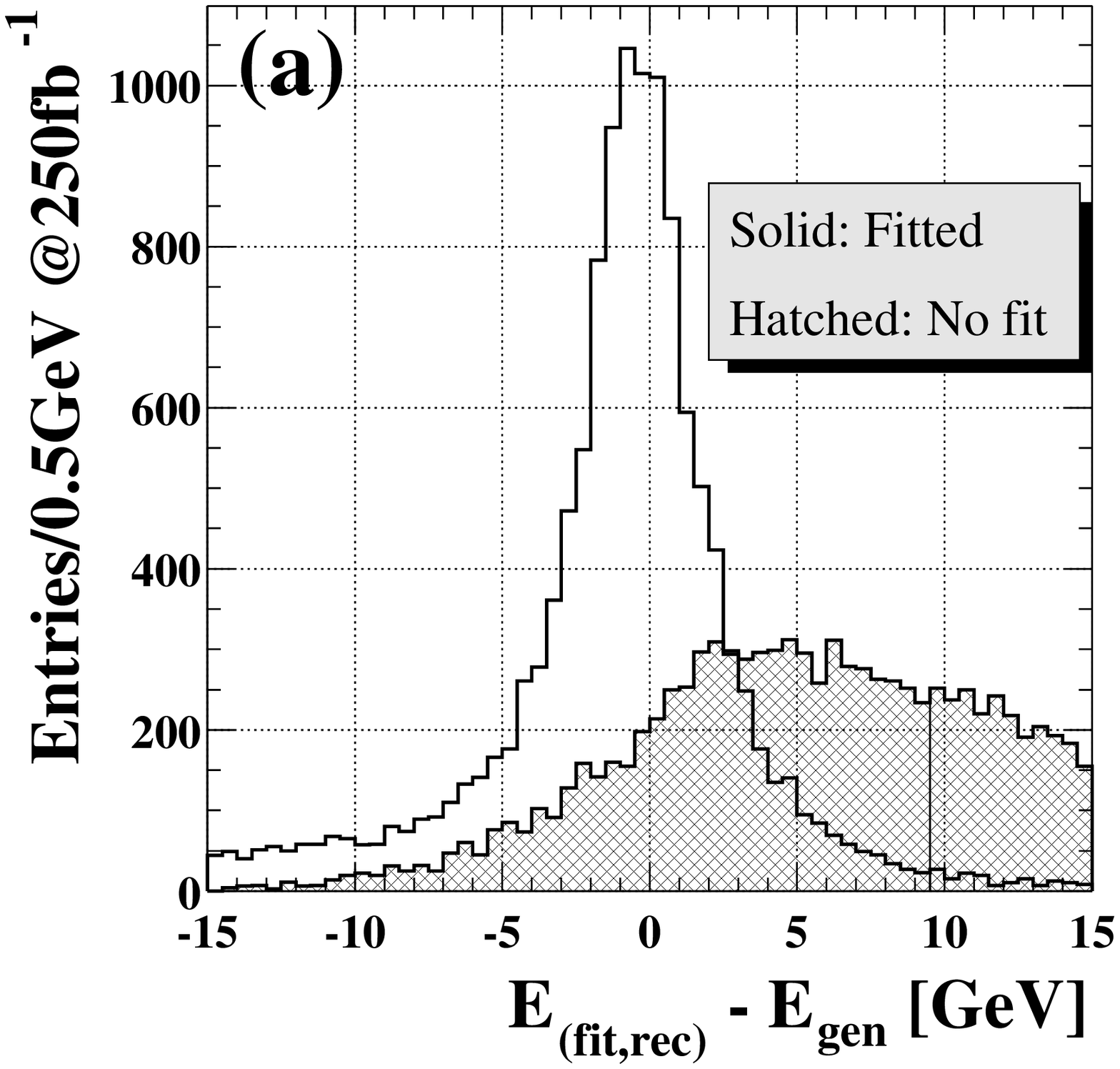}
    \includegraphics[height=5cm,clip]{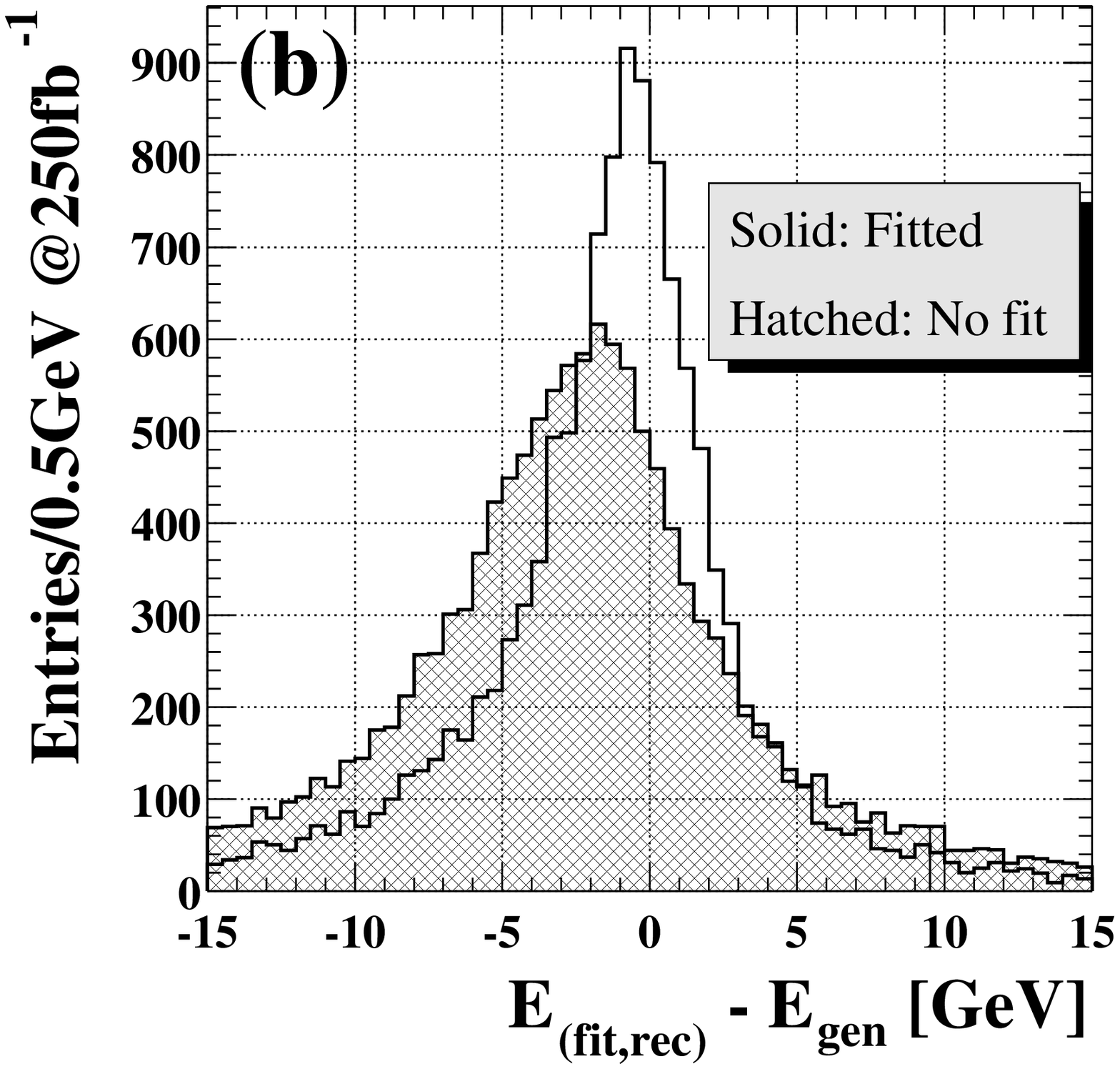}
    \includegraphics[height=5cm,clip]{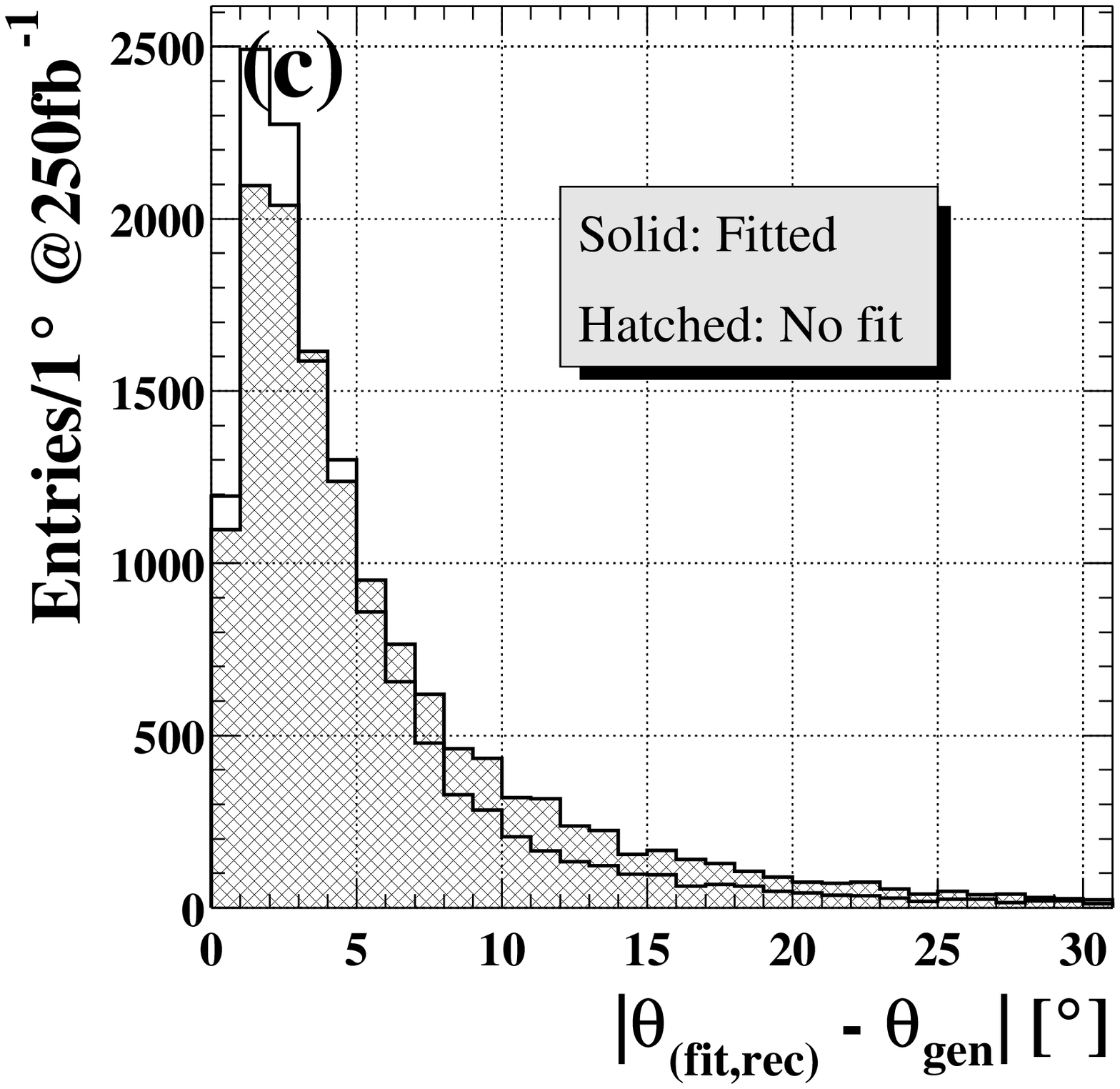}
  \end{center}
  \caption{
	Distributions of the difference of
	the reconstructed and generated energies of
	(a) leptonically-decayed and (b) hadronically-decayed $W$ bosons,
	and (c) distribution of the difference of
	the reconstructed and generated directions of
	the leptonically-decayed $W$,
	before (hatched) and after (solid) the kinematical fit.}
  \label{EresW}
\end{figure}

The improvements in these kinematical variables are reflected
to the improvements in the reconstructed $W$ energies and directions
as shown in Fig.~\ref{EresW}-a) for the energy
of the leptonically-decayed $W$, -b) for the hadronically-decayed $W$,
and -c) for the direction of the leptonically decayed $W$.
We can see dramatic improvements in all of these distributions,
although the improvement in the direction of the hadronically-decayed $W$
is less dramatic.
Both the energy resolution of the leptonically-decayed $W$
and hadronically-decayed $W$ are approximately 2.4~GeV,
the angular resolution of the leptonically-decayed $W$
and hadronically-decayed $W$ are $2.4^\circ$ and $1.7^\circ$, respectively.

\begin{figure}[htbp]
  \begin{center}
    \includegraphics[height=5cm,clip]{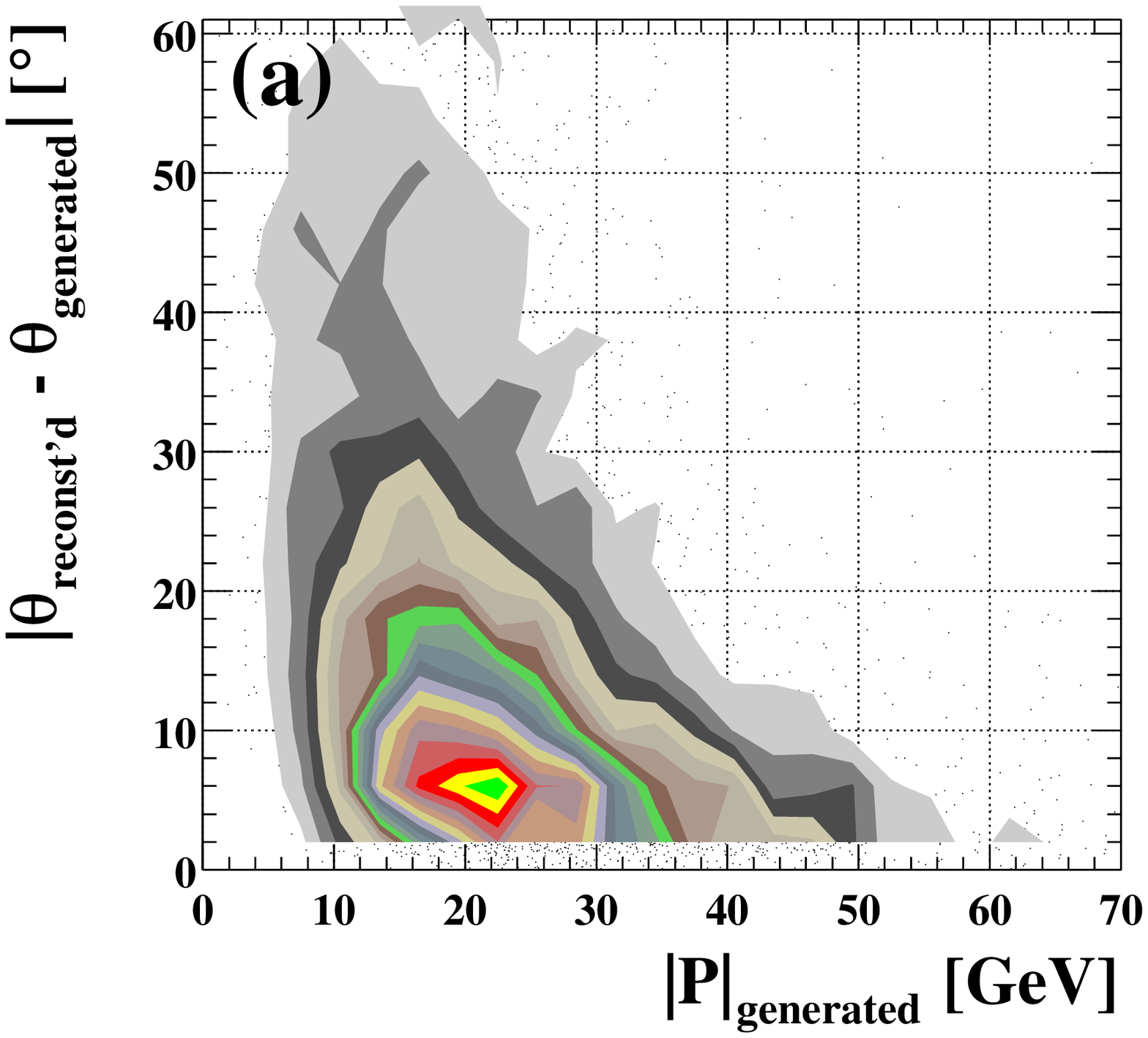}
    \includegraphics[height=5cm,clip]{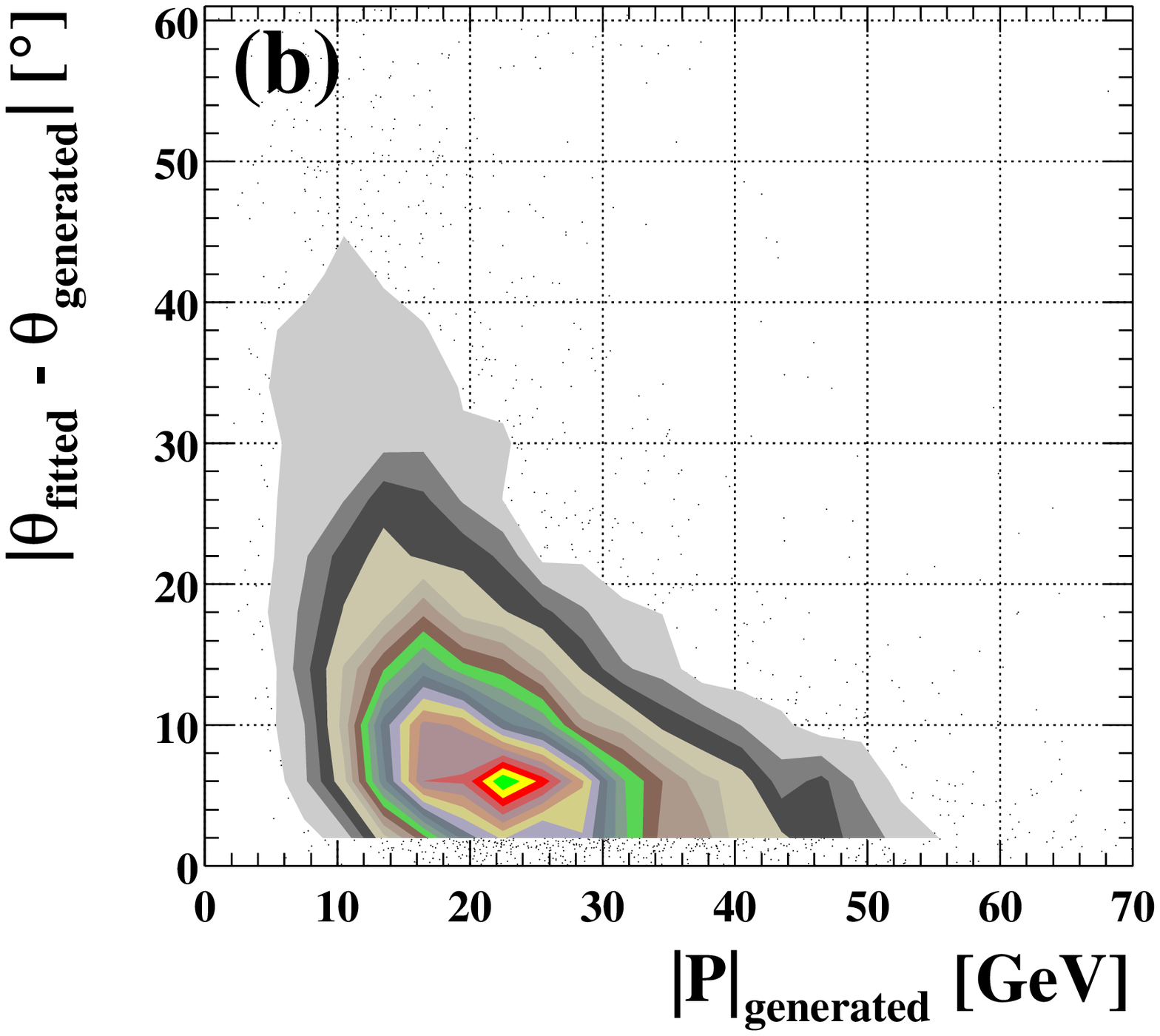}
  \end{center}
  \caption{
	The difference of the reconstructed and generated directions
	of the $t$ or $\bar{t}$ quark plotted against the generated
	top momentum, (a) before and (b) after the kinematical fit.}
  \label{Ares}
\end{figure}

Finally, we will examine the effects of the kinematical fit
on the measurements of the direction and the magnitude
of the top quark momentum.
In Figs.~\ref{Ares}-a) and -b),
the difference of the reconstructed and generated directions
of the $t$ or $\bar{t}$ quark is plotted against the generated top momentum,
before and after the kinematical fit, respectively.
We can see appreciable improvement by the fit.
Nevertheless, since the top quark direction becomes more and more difficult
to measure as the top quark momentum decreases,
the resolution is still somewhat poor in the low momentum region.
The angular resolution is largely determined by the reconstruction
of the $t$ or $\bar{t}$ decayed into 3 jets.
Remember that the resolution improvements were less significant
for the hadronically-decayed $W$,
since the power of the constraints was used up mostly
to recover the momentum information of the direct neutrino
from the leptonically-decayed $W$ and the energy resolution
for jets from the $W$ was left essentially unimproved.
The improvement in the measurements of the top quark direction
is mostly coming from the improvement in the $b$ or $\bar{b}$ jet measurement.
By the same token, the effect of the fit on the measurement
of the magnitude of the top quark momentum is also less dramatic
compared to that on the leptonically-decayed $W$.
The momentum and angular resolutions of the $t$ or $\bar{t}$ quarks
after the fit are approximately 3.0~GeV and $5.5^\circ$, respectively.
\footnote{Since the distributions deviate from Gaussian shapes
substantially off their peaks, these values should be taken as
order of magnitude estimates.}

In the case of the 6-jet mode,
for which there is no direct energetic neutrino from $W$'s,
we can use the power of the constraints to improve
the jet energy measurements.
Consequently, we may expect more significant improvement
in the top quark momentum measurement.

%
\section{A Possible Application}

\begin{figure}[htbp]
  \begin{center}
    \psfrag{cosw}{\hspace{-4mm} \hbox{$\cos \theta_W$}}
    \psfrag{cosl}{$\cos \theta_\ell$}
    (a)
    \begin{minipage}{6.5cm}\centering
      \includegraphics[height=5.5cm,clip]{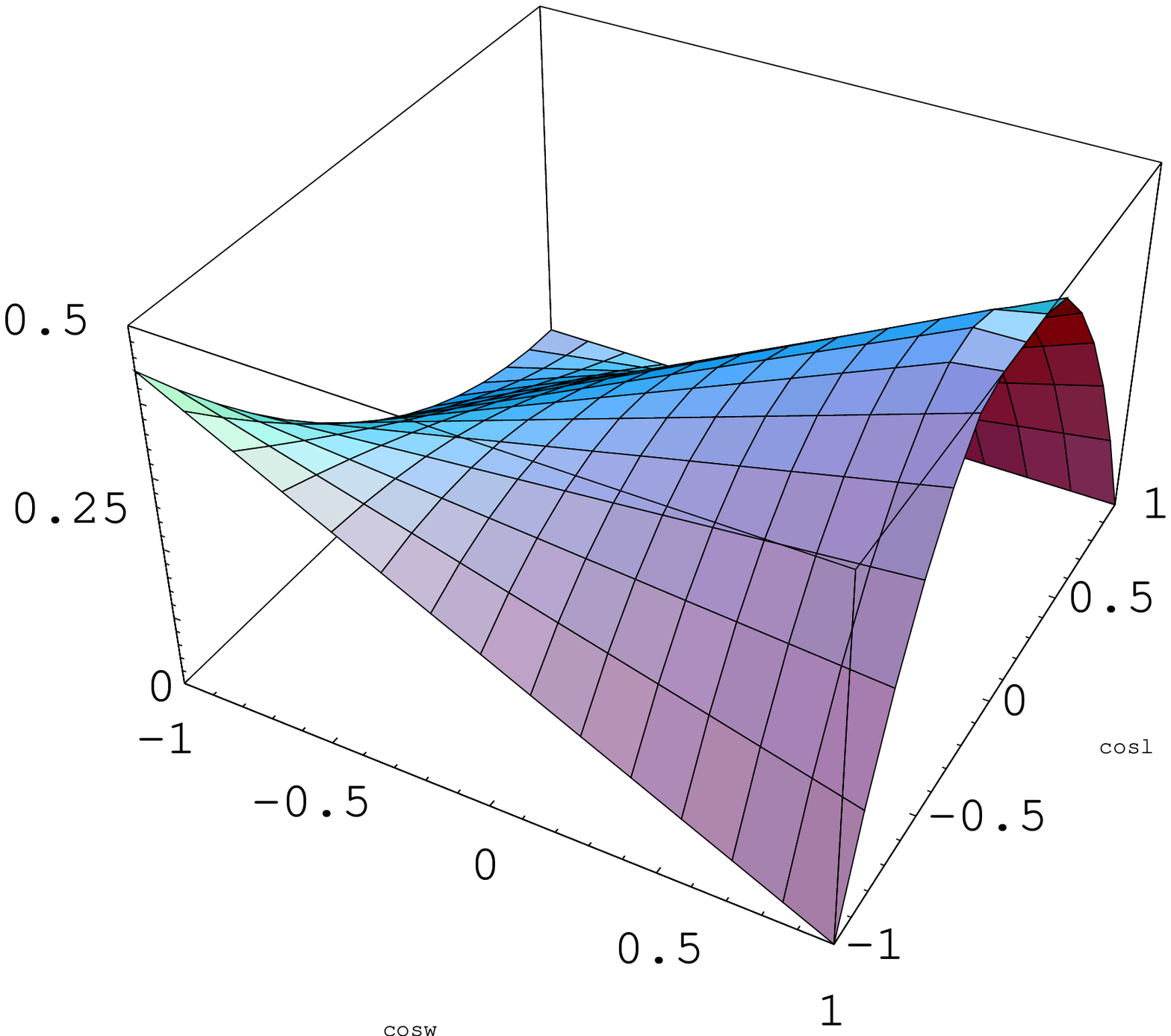}
    \end{minipage}
    \hspace*{1.0cm}
    (b)
    \begin{minipage}{6.5cm}\centering
      \includegraphics[height=5.5cm,clip]{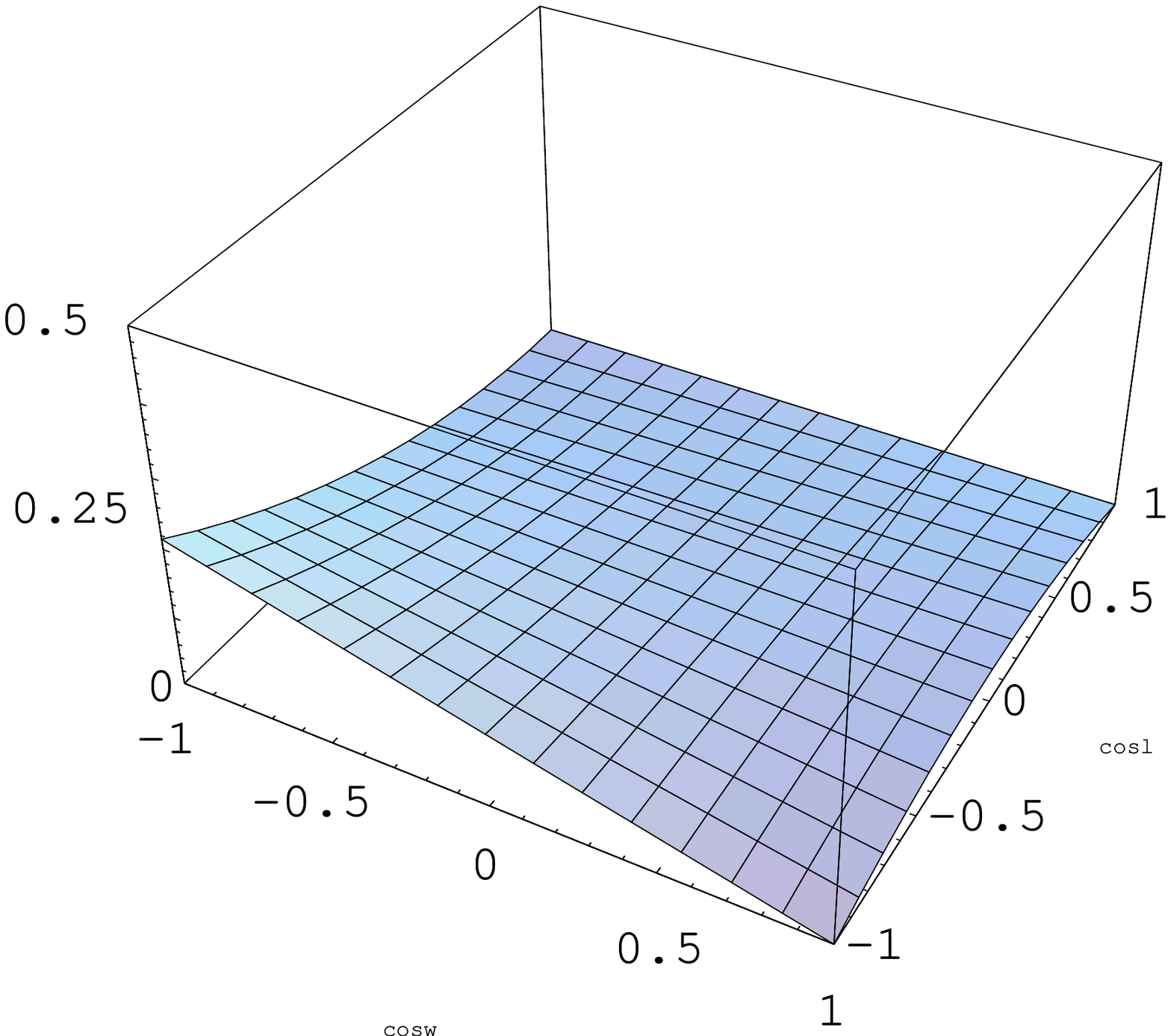}
    \end{minipage}
  \end{center}
  \caption{
	(a) Normalized differential decay width
	${\cal N}^{-1}\, d\Gamma(t_\uparrow \to b\ell\nu)
	/(d \cos \theta_W d \cos \theta_\ell)$
	for $f_2^R=0$.
	(b) Difference of the normalized differential decay widths
	for $f_2^R=0.1$ and for $f_2^R=0$.
	In both figures the differential widths are normalized by
	${\cal N} = \Gamma_t \times {\rm Br}(W\to \ell \nu)$ for $f_2^R=0$.}
  \label{angular-dist}
\end{figure}

We discuss a possible application of our kinematical reconstruction method.
Let us consider measurements of the decay form factors of the top quark
in the $t\bar{t}$ threshold region.
We assume that deviations of the top-decay form factors
from the tree-level SM values are small and consider the deviations
only up to the first order,
i.e.\ we neglect the terms quadratic in the anomalous form factors.
Then the cross sections depend only on two form factors
$f_1^L$ and $f_2^R$ in the limit $m_b \to 0$
although the most general $tbW$ coupling includes
six independent form factors~\cite{Kane:1991bg}:
\bea
\Gamma^\mu_{Wtb} = - \frac{g_W}{\sqrt{2}} \, V_{tb} \,
\bar{u}(p_b) \biggl[
\gamma^\mu \, f_1^L P_L
-
\frac{i\sigma^{\mu\nu}p_{W\nu}}{M_W}
f_2^R P_R
\biggr] u(p_t) ,
\eea
where $P_L = (1-\gamma_5)/2$ and $P_R = (1+\gamma_5)/2$.
At tree level of the SM, $f_1^L =1$ and $f_2^R=0$.
A variation of $f_1^L$ changes only the normalization
of the differential decay width of the top quark,
whereas a variation of $f_2^R$ changes both the normalization and the shape
of the decay distributions.
Thus, we expect that the kinematical reconstruction is useful for
disentanglement of the two form factors
and in particular for the measurement of $f_2^R$.
For simplicity we assume $f_1^L=1$ hereafter.\footnote{
In order to determine $f_1^L$ simultaneously,
we may, for instance, use independent information
from the measurement of the top width~\cite{Fujii:1993mk}.
}
Since transverse $W$ (denoted as $W_T$) is more sensitive
to $f_2^R$ than longitudinal $W$ ($W_L$),
our strategy is to extract $W_T$ using the angular distribution
of $W$ (in the rest frame of $t$) and the angular distribution
of $\ell$ (in the rest frame of $W$).
It is well known that $W_T$ is enhanced in the backward region
$\cos \theta_W \simeq -1$, where the angle $\theta_W$ is measured
from the direction of the top quark spin in the $t$ rest frame.
Also, we may enhance $W_T$ by collecting $\ell$ emitted
in the backward direction $\cos \theta_\ell \simeq -1$,
where the angle $\theta_\ell$ is measured
from the direction of $-\vec{p}_t$ in the $W$ rest frame.
These features are demonstrated in Figs.~\ref{angular-dist}:
We plot\footnote{
We used the helicity amplitudes given in \cite{Kane:1991bg} for calculating
these differential decay widths.
}
(a) the differential decay width for the decay of the top quark
with a definite spin orientation
$d\Gamma(t_\uparrow \to b\ell\nu)/(d \cos \theta_W d \cos \theta_\ell)$
for $f_2^R=0$ and (b) the difference of the differential widths
for $f_2^R=0.1$ and for $f_2^R=0$.
The plots show that we may measure $f_2^R$, for instance,
from the ratio of the numbers of events in the regions
$\cos \theta_W, \cos \theta_\ell < 0$ and
$\cos \theta_W, \cos \theta_\ell > 0$.

\begin{figure}[htbp]
  \begin{center}
    \includegraphics[height=5.5cm,clip]{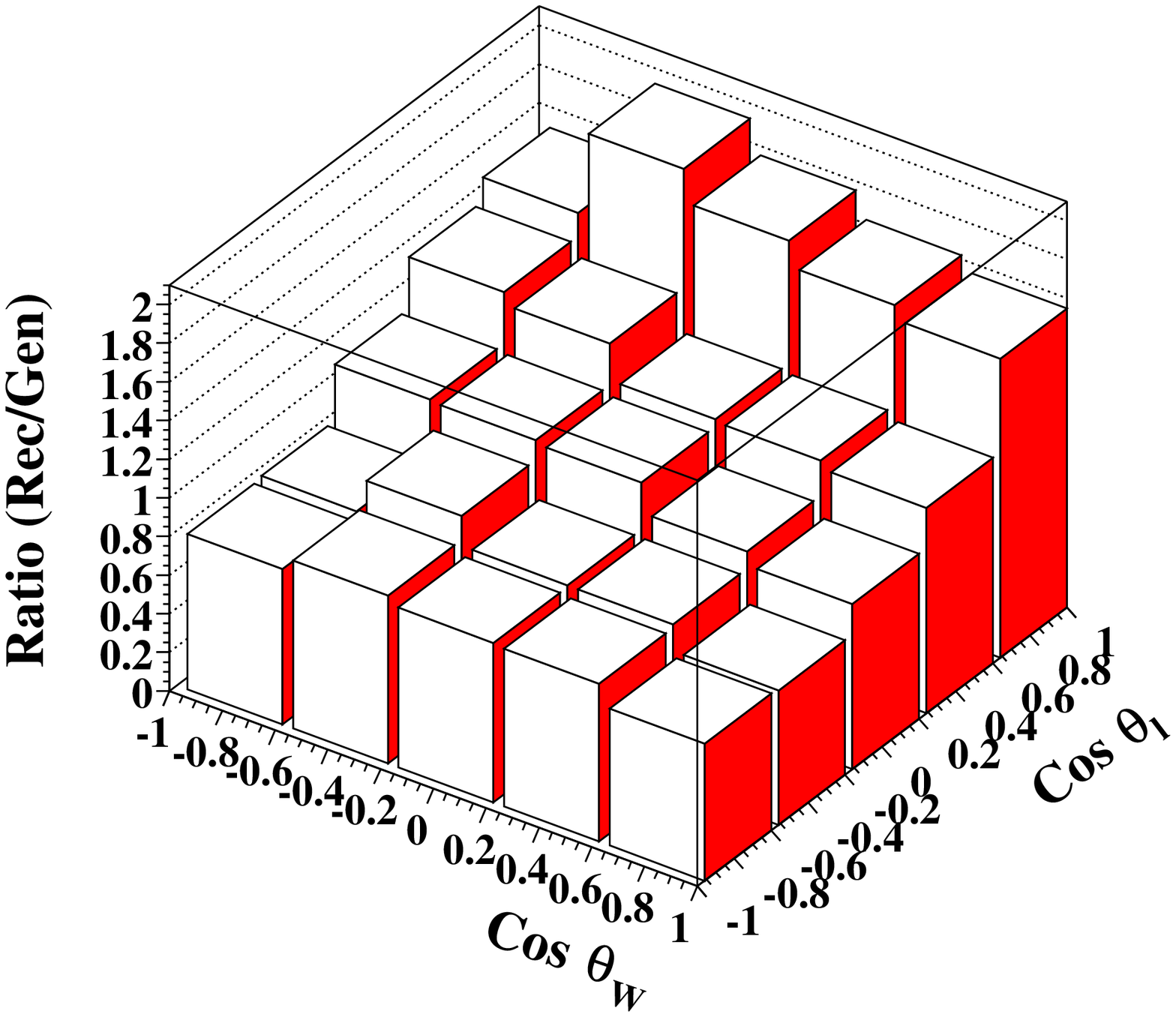}
    \includegraphics[height=5.5cm,clip]{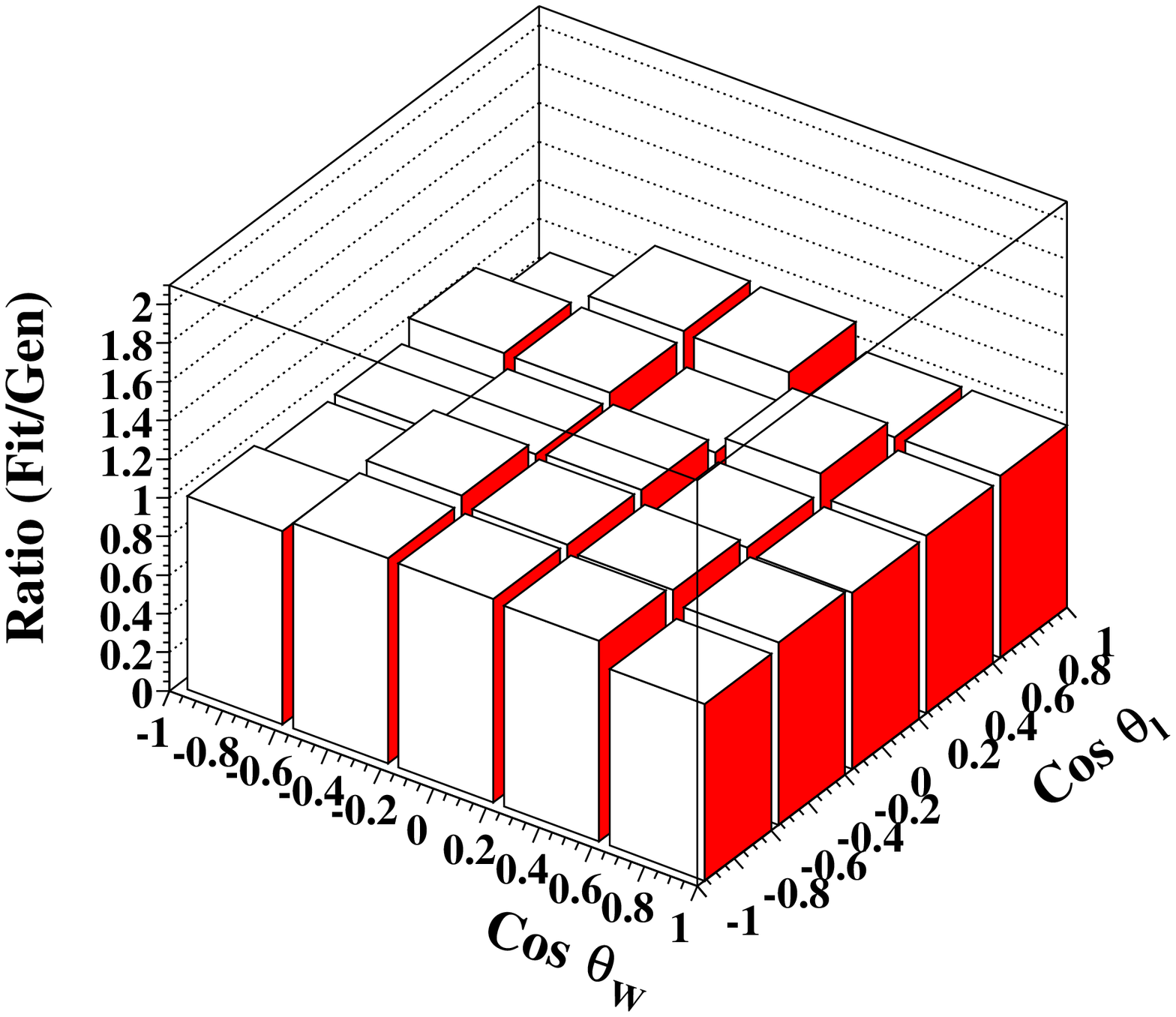}
  \end{center}
  \caption{
	Reconstructed differential decay width distributions corresponding to
	Fig.~\ref{angular-dist}-a) normalized by the generator-level
	distribution (a) before and (b) after the kinematical fit.}
  \label{rec-angular-dist}
\end{figure}

In the last section, we showed the significant improvement
of the reconstruction of the leptonically decayed $W$
due to the kinematical fit.
In order to see how the improvement affects the measurement
of the distribution in question, namely that in Fig.~\ref{angular-dist}-a),
we compare the reconstructed and generated distributions
before and after the kinematical fit,
using the same Monte Carlo sample we used in the previous sections.
Fig.~\ref{rec-angular-dist}-a) and -b) plot,
for the selected $t\bar{t}$ sample, the reconstructed differential decay width
normalized by the corresponding generator level distribution
(a) before and (b) after the kinematical fit.
It is clear from Fig.~\ref{rec-angular-dist}-a) that
the measurement is biased towards high $\cos\theta_{\ell}$,
which is because the energy of the leptonically decayed $W$ tends
to be overestimated so that the lepton from the $W$ is often over-boosted.
Fig.~\ref{rec-angular-dist}-b) demonstrates that the kinematical fit
effectively removed such a measurement bias.
We expect therefore that the kinematical fit will reduce possible
systematic errors in the differential width measurement significantly,
thereby improving sensitivity to $f_2^R$.\footnote{
\samepage
We can extract $f_2^R$ also from the distribution
of $\ell$ energies measured in the laboratory frame
without relying on the reconstruction of its parent $W$ momentum.
The sensitivity of the lepton energy distribution to $f_2^R$ is,
however, estimated to be lower than that of the differential decay width.
}

It is advantageous to investigate decay properties of the top quark
in the $t\bar{t}$ threshold region as compared to the open-top region
$E \gg 2 m_t$ because of several reasons.
First of all, the top quark can be polarized close to 100\%
in the threshold region~\cite{Harlander:1994ac,Harlander:1996vg},
which is a useful tool to sort out various form factors.
This is clear in the above example.
Furthermore, we are almost in the rest frame of the top quark.
In the above example, the top quark is highly polarized in its rest frame.
Hence, the event rate expressed in terms of $\cos\theta_W$
and $\cos\theta_\ell$ is a direct measure of the amplitude-squared,
$|\sum_{i=L,T} {\cal A}(t_\uparrow \to b W_i) \times
{\cal A}(W_i \to \ell\nu)|^2$ (without phase-space Jacobian),
which allows for simple physical interpretations of event shapes.
We also note that we do not gain resolving power
for the decay form factors by raising the c.m. energy.
This is in contrast with the measurements
of the $t\bar{t}$ {\it production} form factors.

%
\section{Summary and Conclusions}

To make maximum use of future $e^+e^-$ linear colliders'
experimental potential,
the top quark reconstruction in the lepton-plus-4-jet mode has been studied
under realistic experimental conditions of $e^{+}e^{-} \to t\bar{t}$ process
near its threshold.
As a new technique to fully reconstruct $t\bar{t}$ final states,
we have developed a kinematical fitting algorithm
which aims to reconstruct the kinematical variables of top quarks
and their offsprings more accurately.

The missing energy carried away by neutrinos from bottom quark decays
has been recovered by the kinematical fitting.
However, the effects of the kinematical fitting
on the top quark momentum are not as dramatic as we wanted.
This is because the top quarks are almost at rest in the threshold region
and therefore their momenta are difficult to measure.
Moreover, in the lepton-plus-4-jet mode many constraints are used up
by recovering the information on the neutrino
from leptonically-decayed $W$ bosons.
On the other hand,
the remarkable improvements of the energy resolution of $b$-jets
and the angular and energy resolutions of leptonically-decayed $W$'s
have been achieved by the kinematical fitting.
These improvements should benefit the form factor measurements in general.
As a possible application,
we considered measurements of decay form factors including $f_2^R$,
on which correct reconstruction of the leptonically-decayed $W$
may have a large impact.

As stated in Sec.~\ref{Sec:introduction},
many theoretical studies on measurements of the top form factors assumed
either the most optimistic case or the most conservative case
with respect to the kinematical reconstruction of event profiles.
Our analysis indicates that both assumptions are not realistic
under actual experimental conditions.
In this respect we emphasize that
the kinematical fit brought often heavily skewed and broad distributions
into nearly Gaussian shapes.
Realistic phenomenological analyses using information
of the decay particles from top quarks will then become possible
by simply Gaussian-smearing parton-level momenta
with the resolutions for the measurements obtained in this study.
To be specific, the resolution for jet energy measurements is
$\sigma_{E_j} \simeq$ 3.5~GeV after the kinematical fit
for both the light quark jets from $W$ boson decays
and the bottom quark jets from $t$ or $\bar{t}$ quarks.
As for the energy resolution for the neutrino coming from
the leptonically decayed $W$ we have $\sigma_{E_{\nu}} \simeq$ 2.5~GeV.
The energy resolutions for both of the leptonically and hadronically
decayed $W$'s then become $\sigma_{E_{W}} \simeq$ 2.4~GeV,
and the angular resolutions for the leptonically decayed $W$
and the neutrino directly coming from it improve to
$2.4^\circ$ and $2.9^\circ$, respectively.
Finally the momentum and angular resolutions
for the the $t$ or $\bar{t}$ quarks
are approximately 3.0~GeV and $5.5^\circ$, respectively.

%
\vskip 0.5cm
\begin{flushleft}
\underline{\bf{Acknowledgements}}
\end{flushleft}
\vskip 0.4cm

The authors wish to thank all the members of the ACFA working group
for useful discussions and comments.
In particular, they are grateful to S.~D.~Rindani for valuable discussions
on strategies for measurements of top quark's possible anomalous couplings,
and A.~Miyamoto for improving JSF (JLC Study Framework)
to incorporate their requests.
This work is partially supported by JSPS-CAS
Scientific Cooperation Program under the Core University System
and the Grant-in-Aid for Scientific Research No.12740130 and No.13135219
from the Japan Society for the Promotion of Science.

%
%

\def\plb#1#2#3{{\it Phys.~Lett.~}{\bf B#1}, #2 (#3)}
\def\prd#1#2#3{{\it Phys.~Rev.~}{\bf D#1}, #2 (#3)}
\def\prl#1#2#3{{\it Phys.~Rev.~Lett.~}{\bf #1}, #2 (#3)}
\def\zpc#1#2#3{{\it Z.~Phys.~}{\bf C#1}, #2 (#3)}

\end{document}